\documentclass[12pt,preprint,referee]{aastex}

\RequirePackage{lineno}
\usepackage{lscape}
\usepackage{gensymb}

\newcommand{\be} {\begin{equation}}

\def\4u0142 {4U\,0142$+$614}
\def\1e1048 {1E\,1048.1$-$5937}
\def\1e1841 {1E\,1841$-$045}
\def\1e1547 {1E\,1547.0$-$5408}
\def\1rxs170849 {1RXS\,J170849.0$-$400910}
\def\xte1810{XTE\,J1810$-$197\,}
\def\cxoj164710{CXOU\,J164710.2$-$455216}
\def\1e2259 {1E\,2259$+$586}
\def\sgr1806{SGR\,1806$-$20}
\def\sgr1900{SGR\,1900$+$14}
\def\sgr1900{SGR\,1935$+$2154}

\def\sgr1627{SGR\,1627$-$41\,}
\def\sgr0501{SGR\,0501$+$4516}
\def\sgr0418{SGR\,0418$+$5729}
\def\sgr1833{SGR\,1833$-$0832}
\def\sgr1801{SGR\,1801$-$21}
\def\sgr1935{SGR\,1935$-$2154}
\def\psrj1622{PSR\,1622$-$4950}
\def\3xmmj185246{3XMM\,J185246.6$+$003317}
\def\swift1822{Swift\,J1822.3$-$1606}
\def\swift1834{Swift\,J1834.9$-$0846}

\newcommand{\fermi}{{\em Fermi}}

\newcommand{\bc}{\begin{center}}
\newcommand{\ec}{\end{center}}
\def\ltsima{$\; \buildrel < \over \sim \;$}
\def\lsim{\lower.5ex\hbox{\ltsima}}
\def\loe{\lower.5ex\hbox{\ltsima}}
\def\gtsima{$\; \buildrel > \over \sim \;$}
\def\gsim{\lower.5ex\hbox{\gtsima}}
\def\goe{\lower.5ex\hbox{\gtsima}}

\def\ltsima{$\; \buildrel < \over \sim \;$}
\def\lsim{\lower.5ex\hbox{\ltsima}}
\def\loe{\lower.5ex\hbox{\ltsima}}
\def\gtsima{$\; \buildrel > \over \sim \;$}
\def\gsim{\lower.5ex\hbox{\gtsima}}
\def\goe{\lower.5ex\hbox{\gtsima}}

\def\ergs {erg\,s$^{-1}$}
\def\ergscm2 { erg\,s$^{-1}$cm$^{-2}$}

\def\cm2 {cm$^{-2}$}

\shortauthors{Fermi-LAT collaboration}
\shorttitle{Fermi-LAT observations of magnetars}

\begin{document}
\title{Gamma-ray upper limits on magnetars with 6 years of {\em Fermi}-LAT observations}

\author{Jian Li\altaffilmark{1}, Nanda Rea\altaffilmark{1,2}, Diego F. Torres\altaffilmark{1,3}, Emma de O$\tilde{n}$a-Wilhelmi\altaffilmark{1}}
\altaffiltext{1}{Institute of Space Sciences (IEEC--CSIC), Campus UAB, Carrer de Magrans s/n, 08193 Barcelona, Spain}
\altaffiltext{2}{Anton Pannekoek Institute for Astronomy, University of Amsterdam, Postbus 94249, NL-1090 GE Amsterdam, the Netherlands}
\altaffiltext{3}{Instituci\'o Catalana de Recerca i Estudis Avan\c{c}ats (ICREA), E-08010 Barcelona, Spain}
\begin{abstract}

We report on the search for gamma-ray emission from 20 magnetars using 6 years of \fermi\, Large Area Telescope (LAT) observations.
No significant evidence for gamma-ray emission from any of the currently-known magnetars is found.
We derived the most stringent upper limits to date on the 0.1--10\, GeV emission of Galactic magnetars, which are estimated between $\sim10^{-12}-10^{-11}$\ergscm2 .
Gamma-ray pulsations were searched for the four magnetars having reliable ephemerides over the observing period, but none were detected.
On the other hand, we also studied the gamma-ray morphology and spectra of seven Supernova Remnants associated or adjacent to the magnetars.
\end{abstract}


\section{INTRODUCTION}
\label{intro}

The magnetars (comprising the Anomalous X-ray Pulsars and Soft Gamma Repeaters; AXPs and SGRs) are a small group of X-ray pulsars (about twenty objects), with spin periods between 0.3 and 12 s.
Their bright X-ray emission ($L_{X}\sim10^{33}-10^{35}$\ergs ) is marginally explained by the {commonly accepted emission models for isolated pulsars} or by accretion from a companion star.
Their inferred magnetic fields assuming the spin-down dipolar loss formula (B = 3.2 $\times$10$^{19} \sqrt{P\dot{P}}$ G, $P$ is the spin period and  $\dot{P}$ is the first derivative of spin period) appear to be in general as high as $B\sim10^{14}-10^{15}$\,G.
Due to these high magnetic fields, the emission of magnetars is thought to be powered by the decay and the instability of their strong fields (Duncan \& Thompson 1992; Thompson \& Duncan 1993, Thompson, Lyutikov, \& Kulkarni 2002).
The powerful X-ray output is usually well modelled by thermal emission from the neutron star hot surface (about 0.2--0.6\,keV), reprocessed in a twisted magnetosphere through resonant cyclotron scattering, a process favored only under these extreme magnetic conditions.
On top of their persistent X-ray emission, magnetars emit very peculiar flares and outbursts on several timescales (from fractions of seconds to years) releasing a large amount of energy ($10^{40}-10^{46}$ erg).
These flares are probably caused by {large-scale} rearrangements of the surface/magnetospheric field, either accompanied or triggered by displacements of the neutron-star crust (somehow analogous to quakes on the stellar surface).
See Mereghetti (2008), Rea \& Esposito (2011), and Olausen \& Kaspi (2014) for recent reviews.
Thanks to \emph{INTEGRAL}, \emph{Suzaku} and  \emph{NuSTAR} we now have good spectra for magnetars in the hard X-ray energy range (Kuiper et al. 2004; Enoto et al. 2010; An et al. 2015).

The gamma-ray emission (0.1--300\,GeV band) from 13 magnetars has been searched using the first 17 months of {\em Fermi}-LAT data (Abdo et al. 2010a).
This search resulted in upper limits and no pulsations were found either.
Taking advantage of about 5 more years of data, in this paper we re-analyze the {\em Fermi}--LAT observations for the previous objects, as well as for the 7 magnetars that were discovered in the meantime.
Furthermore, the better response of the instrument at lower energies brought about by Pass 8 (Atwood et al. 2013) allows us to produce a more stringent search.

In Sec.\,\ref{obs} we report on the {\em Fermi}-LAT analysis procedure, we summarize our results in Sec.\,\ref{results} and \ref{timing}, and provide a
discussion of our findings in Sec.\,\ref{discussion}.

%%%%%%%%%%%%%%%%%%%%%%%%%%%%%%%%%%%
\begin{table*}{}

\centering
\scriptsize
\label{table1}
\caption{{\em Fermi}-LAT upper limits on magnetars as obtained from the likelihood analysis. Fluxes in the different energy ranges
are in units of 10$^{-11}$ erg~cm$^{-2}$s$^{-1}$.}

\begin{tabular}{lcccccc}
\\
\hline\hline                 % inserts double horizontal lines
\\
Source & TS&TS                            &0.1$-$1 GeV & 1$-$10 GeV     & 0.1$-$10 GeV       & {6-years internal list} \\
              &     &(Abdo et al. 2010a)  &   $\Gamma$ (1.5) & $\Gamma$ (3.5)  &  $\Gamma$ (2.5)     &srcs within 3$^{\circ}$ \\
           \\
     \hline
\\
\\
\underline{SGR\,0418$+$5729}                            & 0.0      & 2.3   & $<${0.18}  & $<${0.05 } &$<${0.15}   & 3  \\
\underline{4U\,0142$+$614}                               & 0.0 &3.6     & $<${0.43}  & $<${0.07}  &$<${0.29}     & 3  \\
Swift\,J1822.3$-$1606                           & 0.0 &    & $<$0.88  & $<$0.13  &$<$0.43  &  19  \\
Swift\,J1834.9$-$0846$^\blacktriangledown$$^\star$      & 0.0 &   & $<$1.06 & $<$ 0.99& $<$ 0.45 & 19   \\
\underline{1E\,1048.1$-$5937}                            & 0.0   & 0.0  & $<${0.51}  & $<${0.53}  &$<${1.10}  & 35   \\
XTE\,J1810$-$197                           & 0.0 & 13.1   & $<$1.00  & $<$0.89  &$<$3.65   & 17  \\
\underline{PSR\,J1622$-$4950}                        &  {0.8} &      & $<${1.98}  & $<${0.63}  &$<${1.00} &  26  \\

1E\,1841$-$045 $^\star$                               & 1.0     &7.5  &  $<$1.05 & $<$0.78 & $<$ 2.02 & 23  \\
3XMM\,J185246.6$+$003317$^\star$            & 2.5  &           & $<$1.32  & $<$0.49  &$<$1.88  & 23   \\

\underline{1E\,2259$+$586}$^\star$                       &  {2.8}   & 15.6 & $<${0.21}& $<${0.12}& $<${0.13} & 13 \\

SGR\,1806$-$20                          & 2.9 & 2.8 & $<$2.38  & $<$0.50  &$<$0.67  &  15  \\

SGR\,1935$+$2154                        & 3.5 &             & $<$0.11          & $<$0.22        &$<$0.17       & 6 \\
\underline{SGR\,0501$+$4516}$^\star$                             &  {3.8}   & 16.3 & $<${0.51}  & $<${0.08}  &$<${0.40}       & 4  \\
\underline{1E\,1547.0$-$5408}                           & {4.2} & 36.2 & $<${1.40} & $<${0.90}  &$<${2.84}  & 14  \\

SGR\,1900$+$14$^\star$                           & 4.5 &0.0  & $<$ 0.63  & $<$0.49  &$<$ 2.07  & 13  \\

\underline{CXOU\,J164710.2$-$455216}                 &{6.0}&  & $<${0.96}  & $<${0.45}  &$<${0.89}& 27  \\
\underline{1RXS\,J170849.0$-$400910}                 & {6.6} & 32.1 & $<${2.14}  & $<${0.96}  &$<${2.51}  & 15  \\
SGR\,1833$-$0832   $^\blacktriangledown$$^\star$                       & 18.9 && $<$2.75  & $<$1.29  &$<$0.87  & 19   \\

\underline{SGR\,1627$-$41}$^\star$                           & {23.8}  & 36.0 & $<${2.87}  & $<${1.75}  & $<${3.34} & 29   \\

CXOU\,J171405.7$-$381031$\S$          & 51.0$\dag$  &           & $<$0.27  & $<$0.53  &$<$0.80  & 11   \\

\tablecomments  {Properties of the magnetars studied in this work, ordered by the measured TS values derived from the binned analysis.
{The magnetars underlined have zenith angle cut at 90$\degree$ (see section \ref{obs} for detail).}
The GeV upper limits are reported at {95\% confidence level} (see Section \ref{obs} for details).
$\blacktriangledown$: SGR\,1833$-$0832 and Swift\,J1834.9$-$0846 are included in the same spatial model.
$\S$: the upper limits of CXOU\,J171405.7$-$381031 are estimated with photon index fixed at 1.71.
$\star$: magnetars with associated or close-by SNRs, the TS value are the residual after modelling the {extended} SNRs.
$\dag$:{The high TS value may come from the possible
%point-like
gamma-ray emitting SNR CTB 37B. See section \ref{CXOU_subsection} for more detail.}}
\end{tabular}
\end{table*}

%%%%%%%%%%%%%%%%%%%%%%%%%%%%%%%%%%%

\section{OBSERVATIONS AND DATA ANALYSIS}
\label{obs}
The \emph{Fermi}-LAT data included in this paper covered 74 months from 2008 August 4 to 2014 October 1, which greatly extended the 17 months data coverage of Abdo et al. (2010a).
The analysis of \emph{Fermi}-LAT data was performed using the \emph{Fermi} Science Tools\footnote{\url{http://fermi.gsfc.nasa.gov/ssc/}}, 09-34-01 release.
{Events from the Pass 8 source class were selected.}
{The P8 SOURCE V5 instruments response functions (IRFs) were used for the analysis\footnote{The Science Tools and IRFs used here are internal pre-release versions of the Pass 8 data analysis.  Our results did not change substantially with the final release versions.}.}
All gamma-ray photons within an energy range of 0.1--300 GeV and within a circular region of interest (ROI) of 10$\degree$ radius centered on each source analyzed were used for analysis.
To reject contaminating gamma rays from the Earth's limb, we selected events with zenith angle $<$ 100$\degree$, {or $<$ 90$\degree$ for the sources with high possibility of contamination (high-latitude sources with delc.$>$ 45$\degree$, underlined in Table 1).}
The gamma-ray flux and spectral results presented in this work were calculated by performing a binned maximum likelihood fit using the Science Tool \emph{gtlike}.
{The binned maximum likelihood analysis was carried out on a 14$\fdg$1$\times$14$\fdg$1 square region with a pixel size of 0$\fdg$1 centered on each magnetar.}
{30 bins in 100 MeV--300 GeV were used in the analysis.}
The spectral-spatial model constructed to perform the likelihood analysis includes Galactic and isotropic diffuse emission components{ (``gll\_iem\_v06.fits", Acero et al. 2016b, and ``iso\_P8R2\_SOURCE\_V6\_v06.txt", respectively\footnote{\url{http://fermi.gsfc.nasa.gov/ssc/data/access/lat/BackgroundModels.html}})} as well as known gamma-ray sources within 15$\degree$ from the magnetars, following the internal 6-years source list which was based on ``Pass 8'' \emph{Fermi}-LAT data covering from 2008 August 4 to 2014 August 4.
The spectral parameters were fixed to the catalog values, except for the sources within 3$\degree$ from the magnetars.
For these latter sources, all the spectral parameters were left free.

Magnetars in Table 1 were modelled with a simple power law with all spectral parameters allowed to vary.
The maximum likelihood test statistic (TS) was employed to evaluate the significance of the gamma-ray fluxes coming from the magnetars.
The Test Statistic is defined as TS=$-2 \ln (L_{max, 0}/L_{max, 1}) $, where $L_{max, 0}$ is the maximum likelihood value for a model without an additional source (the ``null hypothesis") and $L_{max, 1}$ is the maximum likelihood value for a model with the additional source at a specified location.
A larger TS indicates that the null hypothesis is incorrect (i.e., a source is detected).
The square root of the TS is approximately equal to the detection significance of a given source.
We set the detection threshold to be TS$>$25.
The 95$\%$ flux upper limits for magnetars are calculated following Helene's method (Helene 1983) assuming a {power-law} spectral shape.
In the 0.1--10 GeV energy range, we fixed the photon index at 2.5, as in Abdo et al. (2010a).
In 0.1--1 GeV and 1--10 GeV ranges, the upper limits are evaluated using a photon index fixed at 1.5 and 3.5 respectively, mimicking a spectral cutoff at $\sim$ 1GeV, as is common in pulsar spectra.
{Apart from the photon index, other spectral parameters in the upper limits calculation are the same as in the \emph{gtlike} analysis.}
For extended source analysis, we followed the method as described in Lande et al. (2012).
We assumed the source to be spatially extended with a symmetric disk model and fitted its position and extension with the \textit{Pointlike} analysis package (Kerr, 2011) between 0.1 and 300 GeV.
Following the method in Lande et al. (2012), the TS value of the putative extension is defined as TS$_{ext}$=2($\ln L_{disk}$-$\ln L_{point}$), in which $L_{disk}$ and   $L_{point}$ were the \textit{gtlike} global log likelihood of the point source and extended source hypothesis, respectively.
We set the threshold for claiming the source to be spatially extended as TS$_{ext}>$16 corresponding to a significance of $\sim$ 4 $\sigma$.
TS maps shown in this paper are produced with the Pointlike analysis tool.
{In most cases, the \emph{Fermi}-LAT data could be well modelled by the {sources from the internal 6-years source list}.
In case of excesses in the residual TS map, additional point sources were added following a similar method to that described in Caliandro et al. (2013), section 3.3.3.}

%%%%%%%%%%%%%%%%%%%%%%%%%%%%%%%%%%%
\begin{center}
\begin{figure*}
\centering
\vbox{
\hbox{
\centering
\includegraphics[scale=0.28]{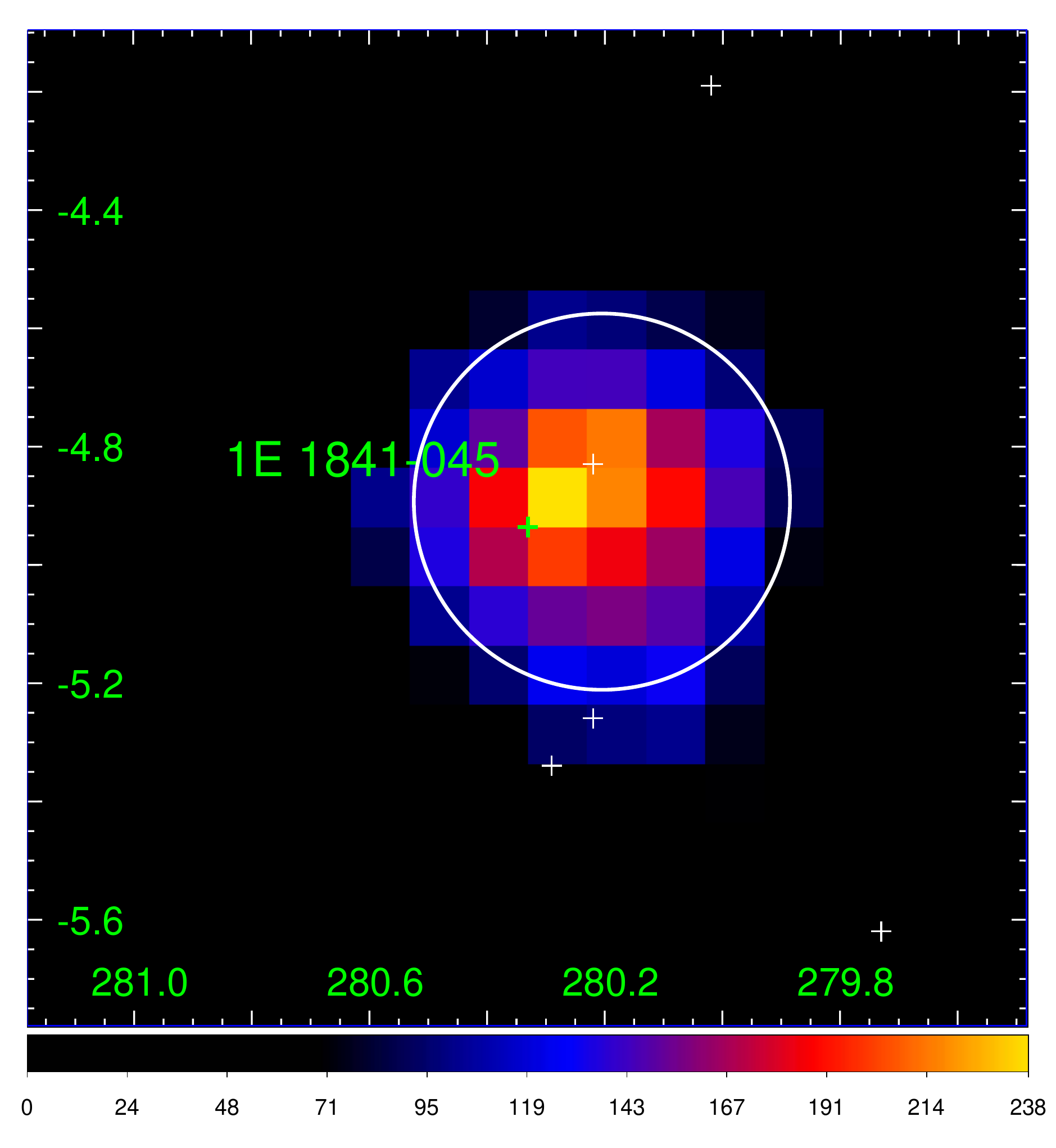}
\includegraphics[scale=0.28]{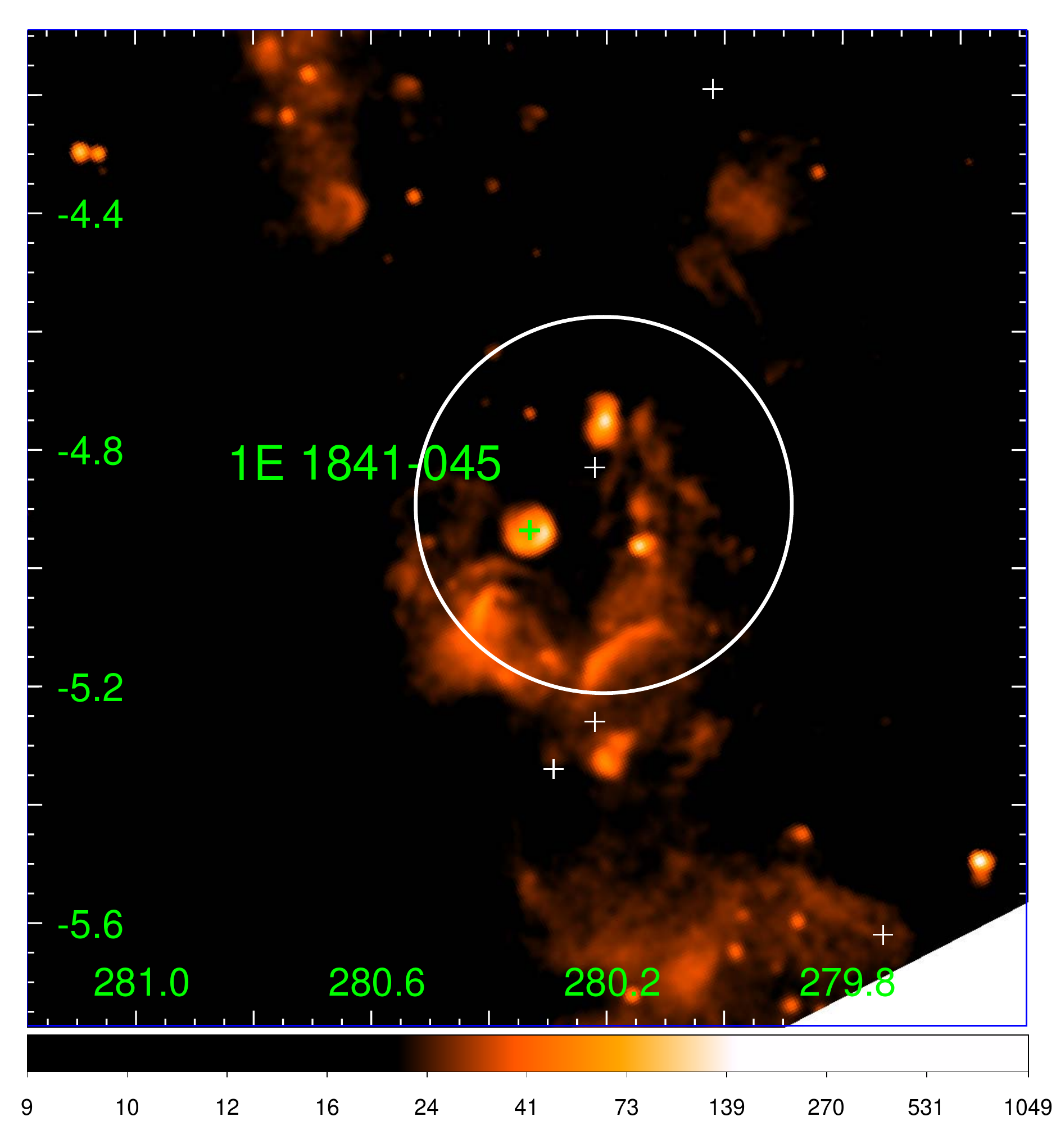}
\includegraphics[scale=0.28]{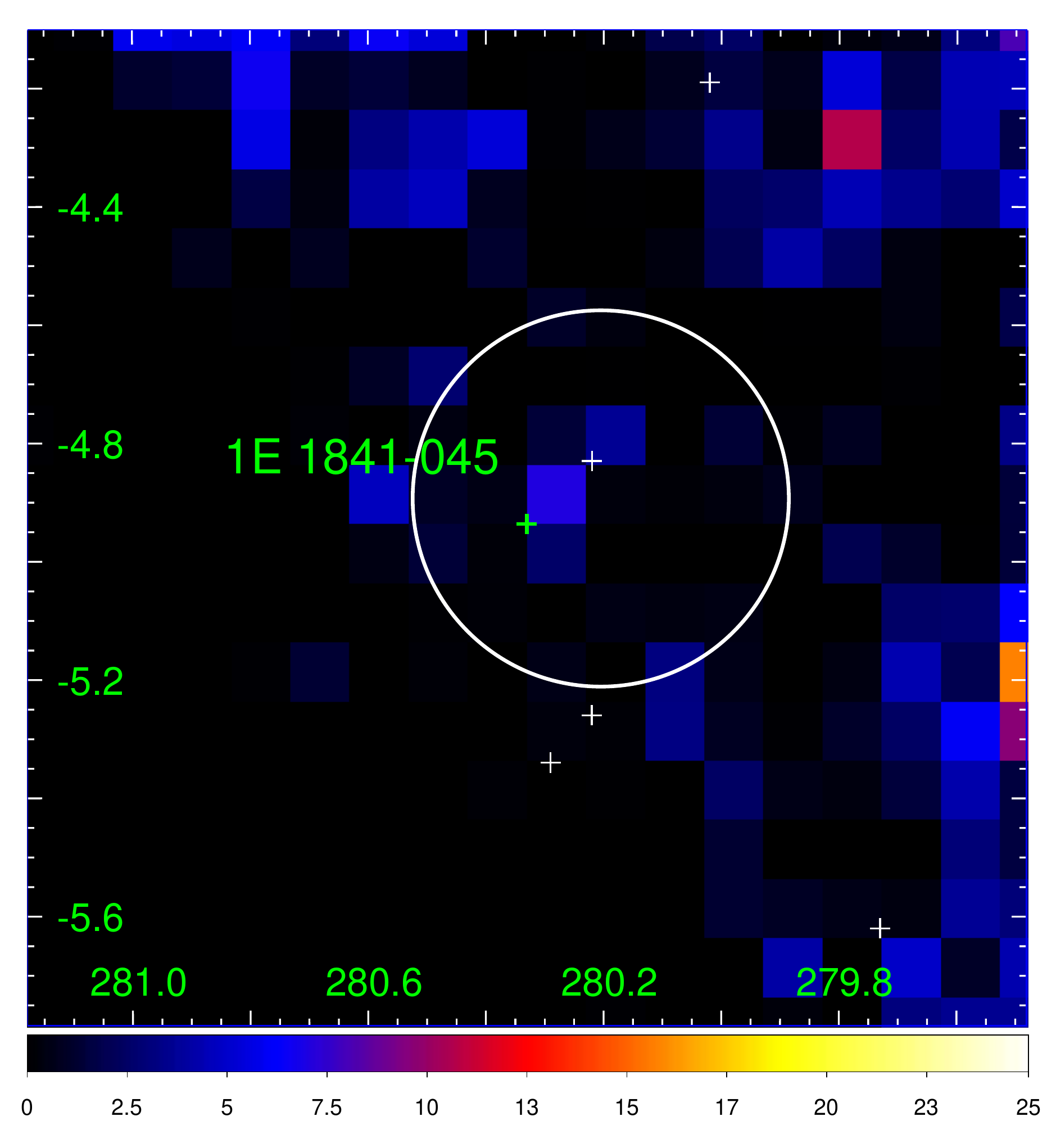}}
\hbox{
\includegraphics[scale=0.28]{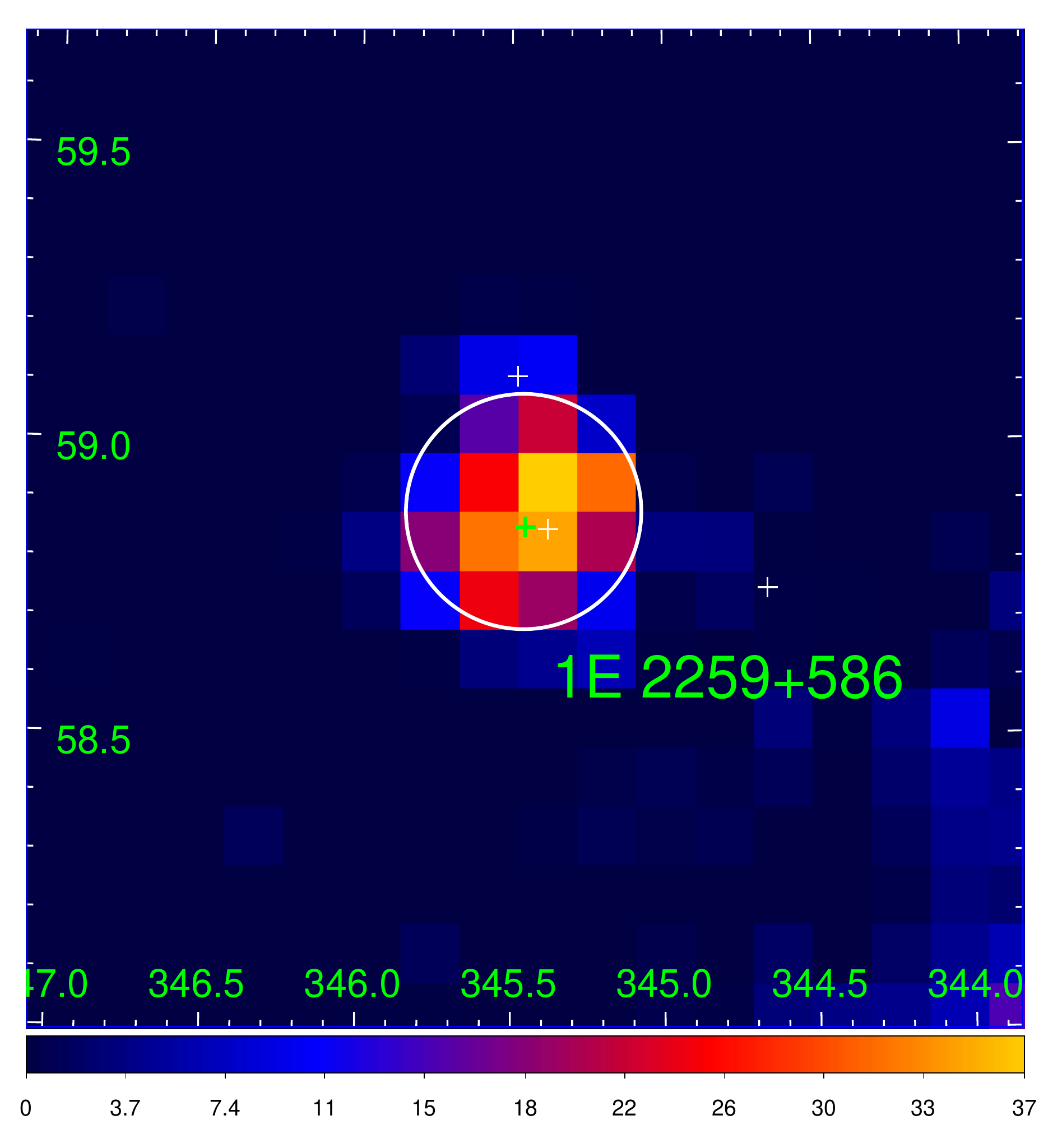}
\includegraphics[scale=0.28]{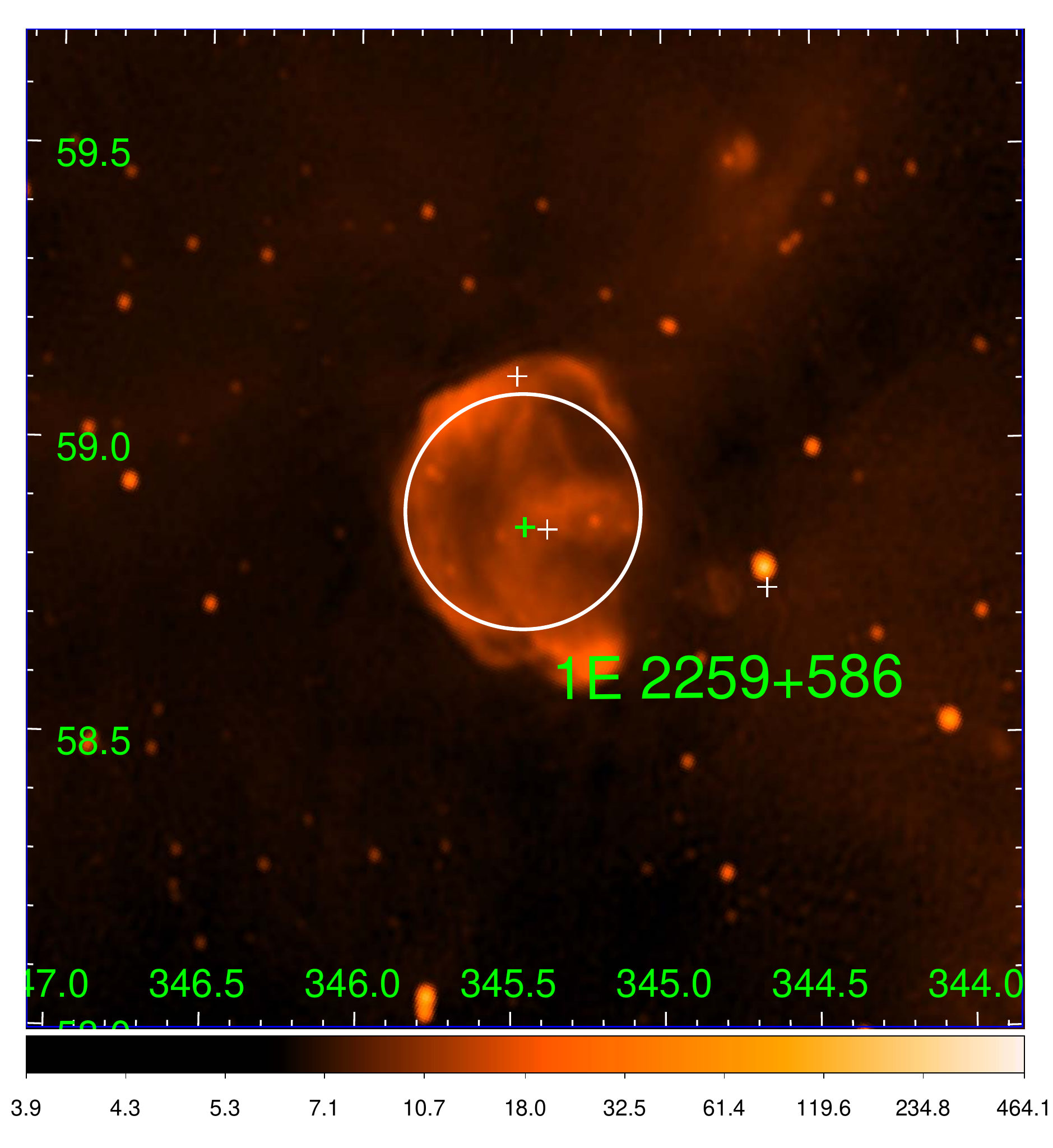}
\includegraphics[scale=0.28]{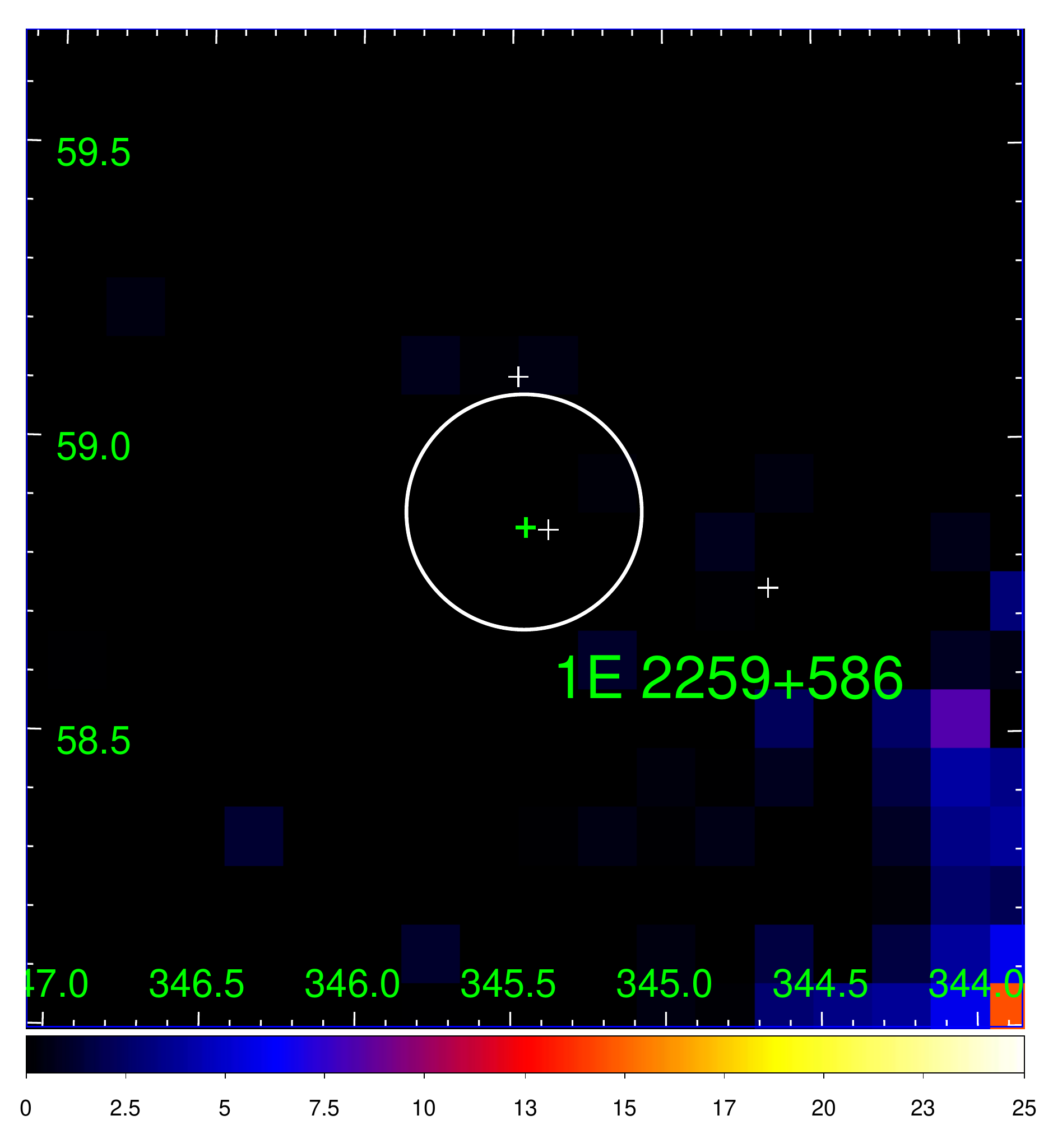}}
\hbox{
\includegraphics[scale=0.28]{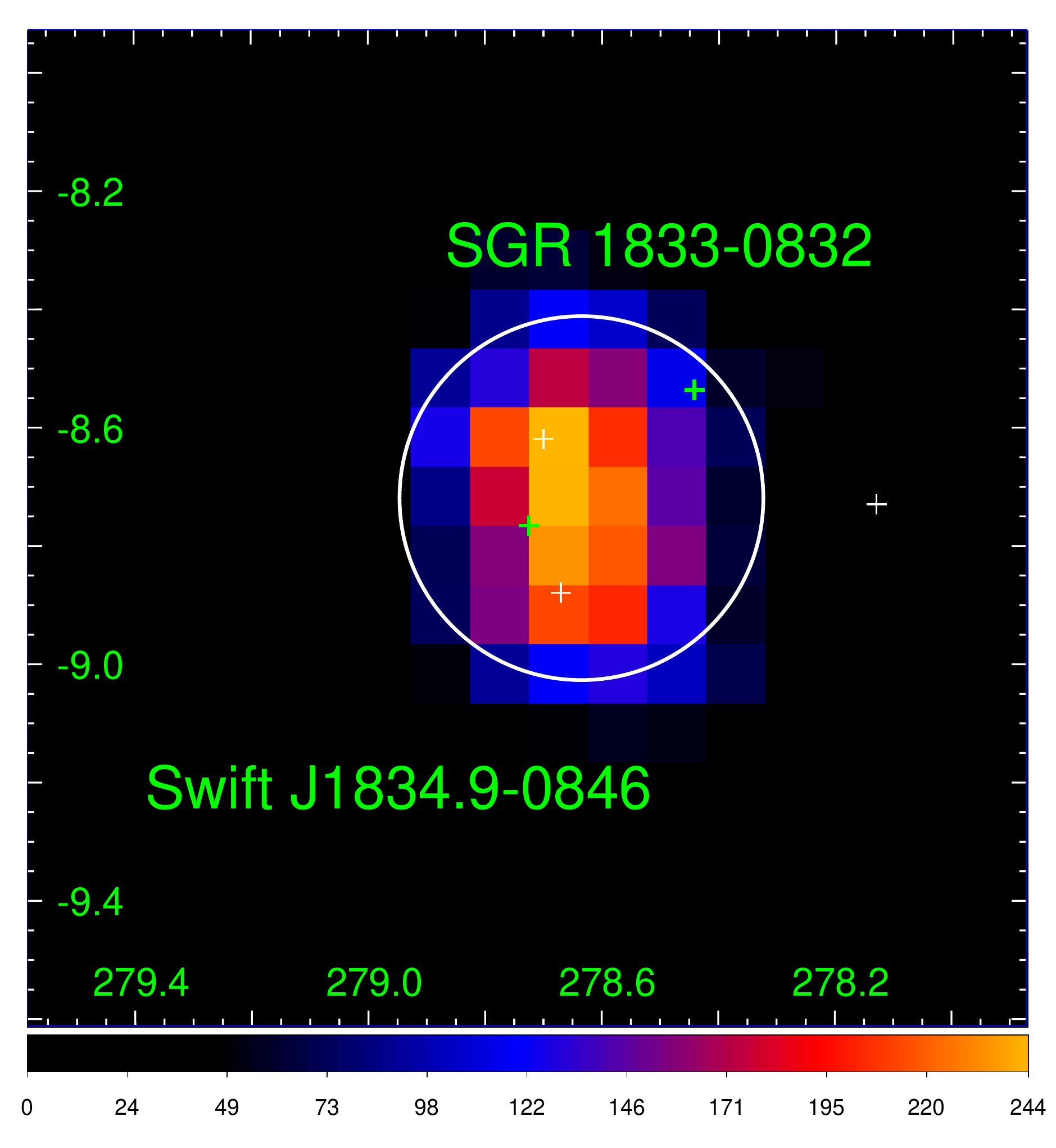}
\includegraphics[scale=0.28]{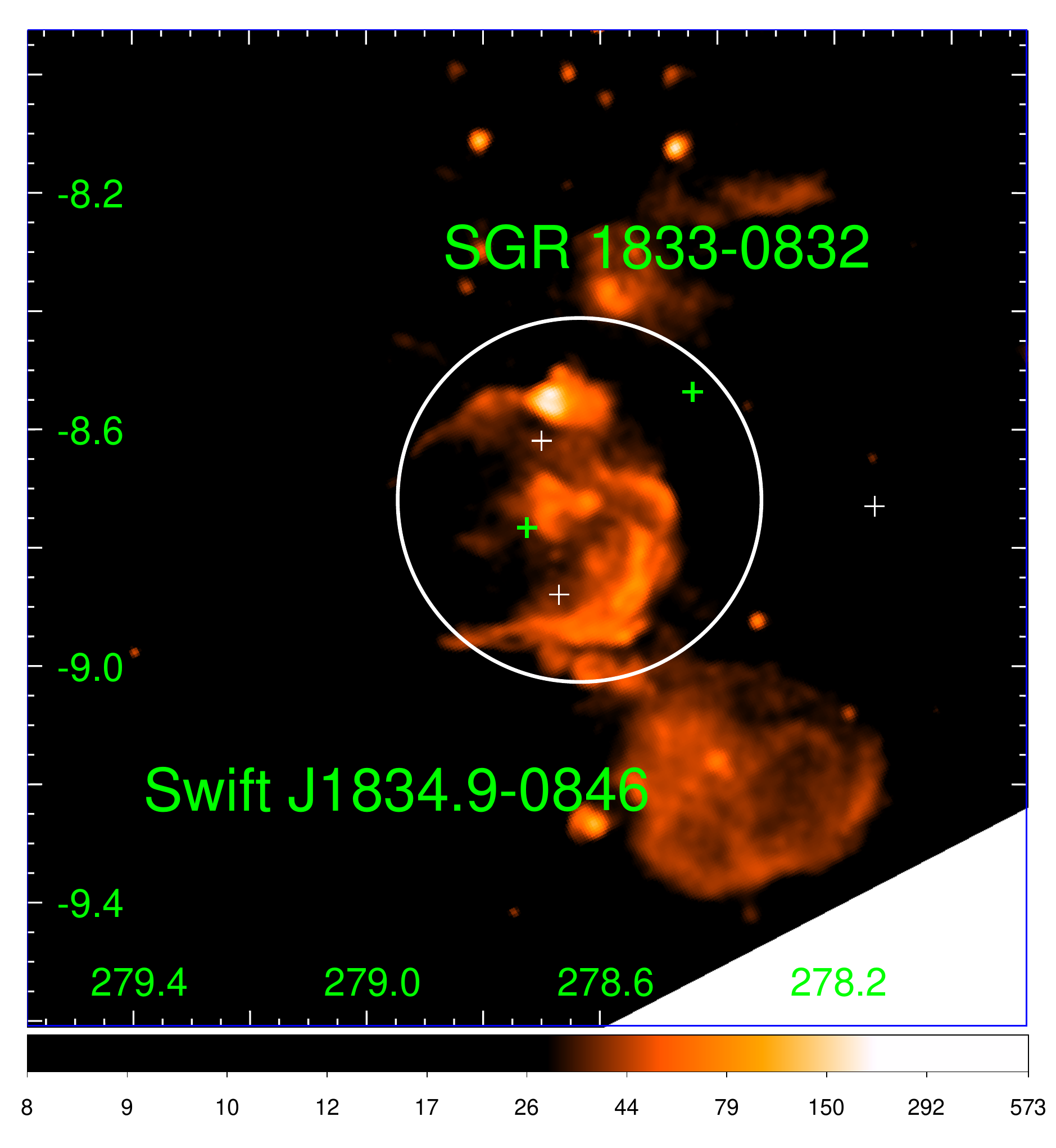}
\includegraphics[scale=0.28]{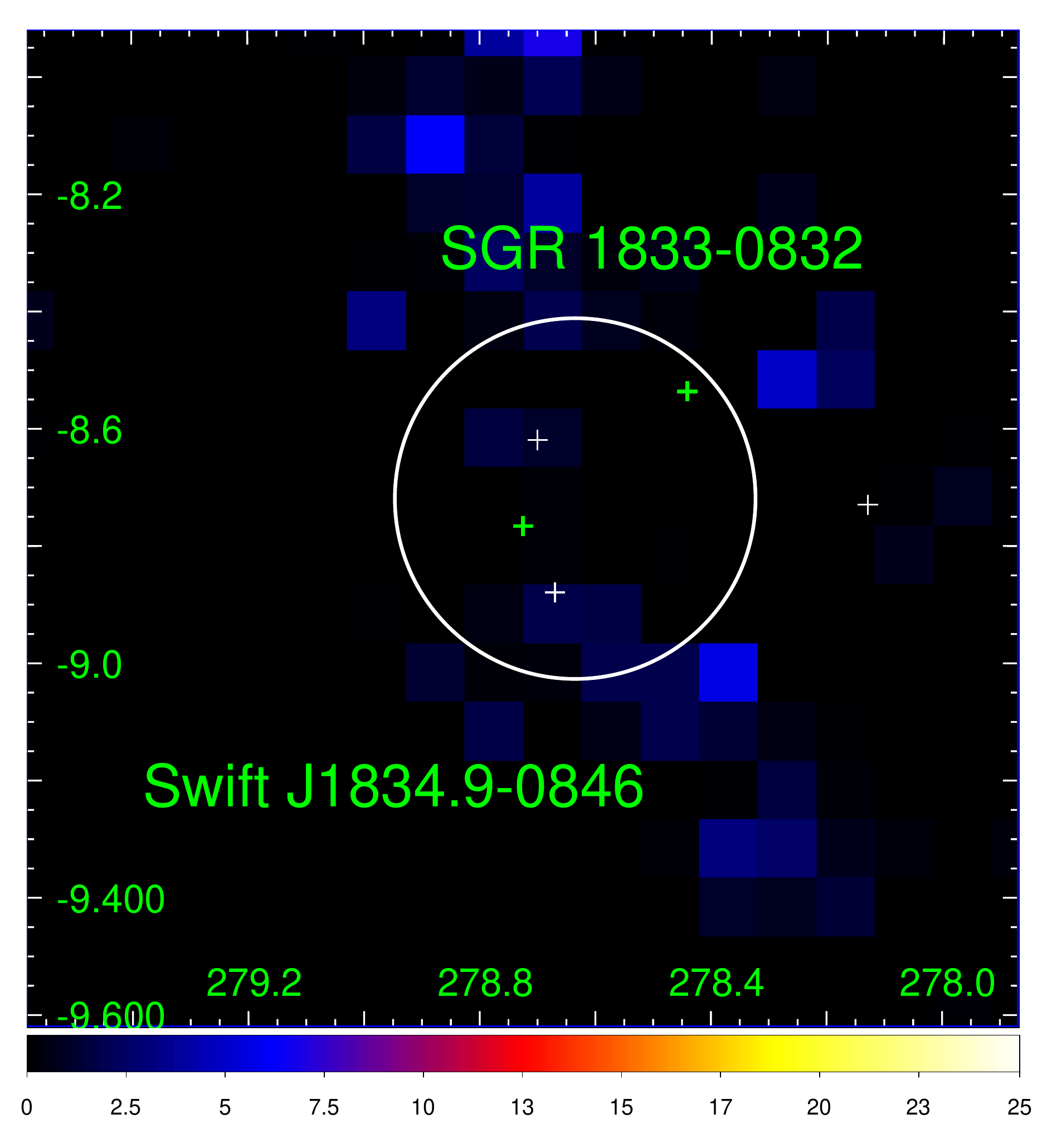}}}
\caption{\fermi-LAT fields of 1E\,1841$-$045, 1E\,2259$+$586, Swift\,J1834.9$-$0846, and SGR\,1833$-$0832.
The left column shows the relative TS maps in 0.1-300 GeV.
The middle columns are the radio map of the same region from CGPS for 1E\,2259$+$586, and VGPS for the other sources.
The right column shows the residual TS maps after modelling the gamma-ray emission of the relative SNRs, namely SNR Kes 73, CTB 109, and W41 for 1E\,1841$-$045, 1E\,2259$+$586, and Swift\,J1834.9$-$0846, respectively.
The X-axis and Y-axis are {R.A. and decl.} referenced to J2000.
The magnetars are shown with a green cross while {other sources considered in the model} are shown with a white cross.
The {best-fit} disk template are shown with a white circle.
Each source is zoomed on at  the same scale and the size of disks can be directly compared.
The spatial models used to produce TS maps are described in the text.}
\label{1E1841}
\label{1E2259}
\label{Swift1834}
\end{figure*}
\end{center}
%%%%%%%%%%%%%%%%%%%%%%%%%%%%%%%%%%

\section{Results}
\label{results}

We studied the {\em Fermi}--LAT {observations} of all confirmed magnetars from the McGill Online Magnetar Catalog\footnote{\url{http://www.physics.mcgill.ca/$\sim$pulsar/magnetar/main.html}} (Olausen \& Kaspi, 2014), except for the extra-galactic ones and Galactic center magnetars due to the difficulty of resolving them from their respective environments (see Abdo et al. 2010b, 2010c).
Those are: SGR 0526$-$66 in the Large Magellanic Cloud, CXOU J010043.1$-$721134 in the Small Magellanic Cloud and  SGR 1745$-$2900 at the Galactic center. We list in Table 1 the results for the 20 selected magnetars.

Most of the magnetars are clearly undetected, having TS values $<25$ in the \textit{gtlike} analysis. This is the case of: SGR\,0418$+$5729, 4U\,0142$+$614, Swift\,J1822.3$-$1606, PSR\,J1622$-$4950, 1E\,1048.1$-$5937, XTE\,J1810$-$197, SGR\, 1806$-$20, SGR\,1935$+$2154, CXOU J164710.2$-$455216, and {1E\,1547.0$-$5408}.
No gamma-ray source is associated with these 10 magnetars in the TS maps.
Upper limits are evaluated for the 10 magnetars.

The uncertainties of the {\em Fermi}--LAT effective area and of the Galactic diffuse emission are the two main sources of systematics that can affect the evaluation of the upper limits {to a different extent on individual sources.}
These systematic effects are estimated by repeating the upper limit analysis using modified IRFs that bracket the effective areas and changing the normalization of the Galactic diffuse model artificially by $\pm$6\%, similar to Abdo et al. (2010a), for all magnetars considered in this paper\footnote{An alternative method of estimating systematics is reported in Acero et al. (2016a).}.
Energy dispersion is not considered in the data analysis, which may be important below 100 MeV, but not expected to produce significant changes in the {100 MeV--300 GeV} energy range considered in this work.
The results of this analysis are reported in Table 1.

Below we comment in a case by case basis on all magnetar locations that resulted in a high TS value since they are either associated with or adjacent to a {\em Fermi}-LAT detected SNR, or located in a crowded region.

%%%%%%%%%%%%%%%%%%%%%%%%%%%%%%%%%%%
\begin{center}
\begin{figure*}
\centering
\vbox{
\hbox{
\includegraphics[scale=0.27]{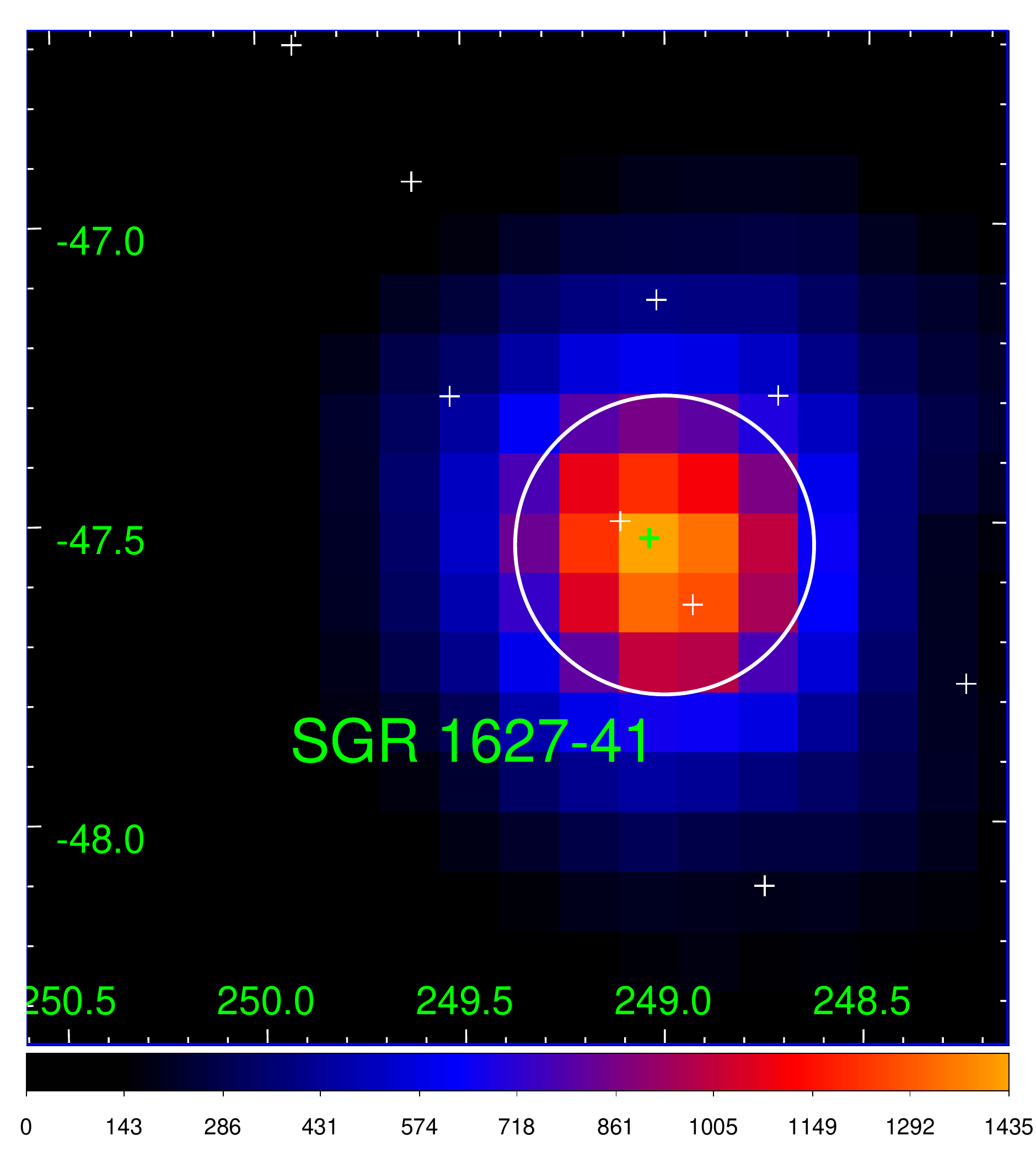}
\includegraphics[scale=0.27]{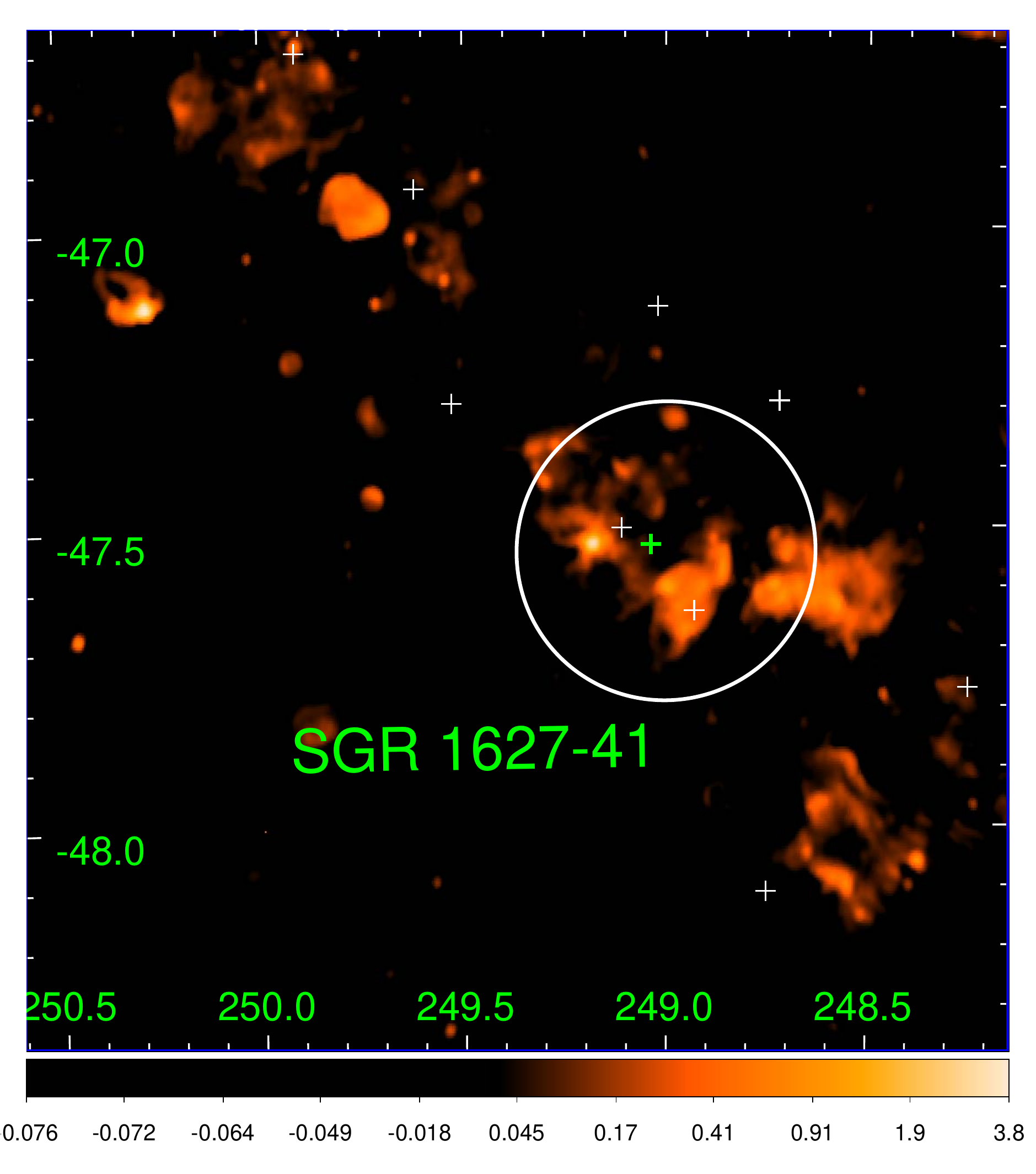}
\includegraphics[scale=0.27]{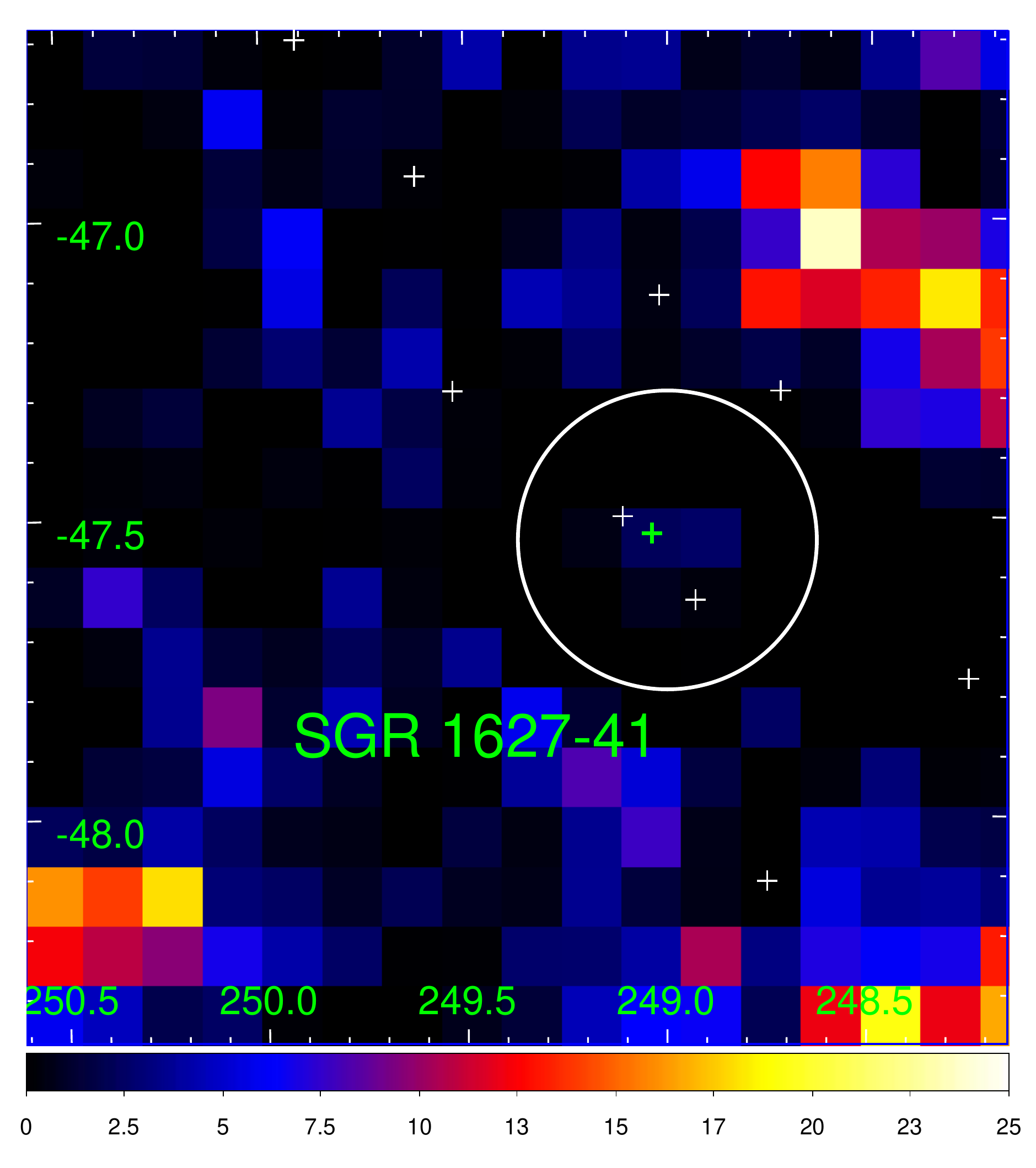}}
\hbox{
\includegraphics[scale=0.27]{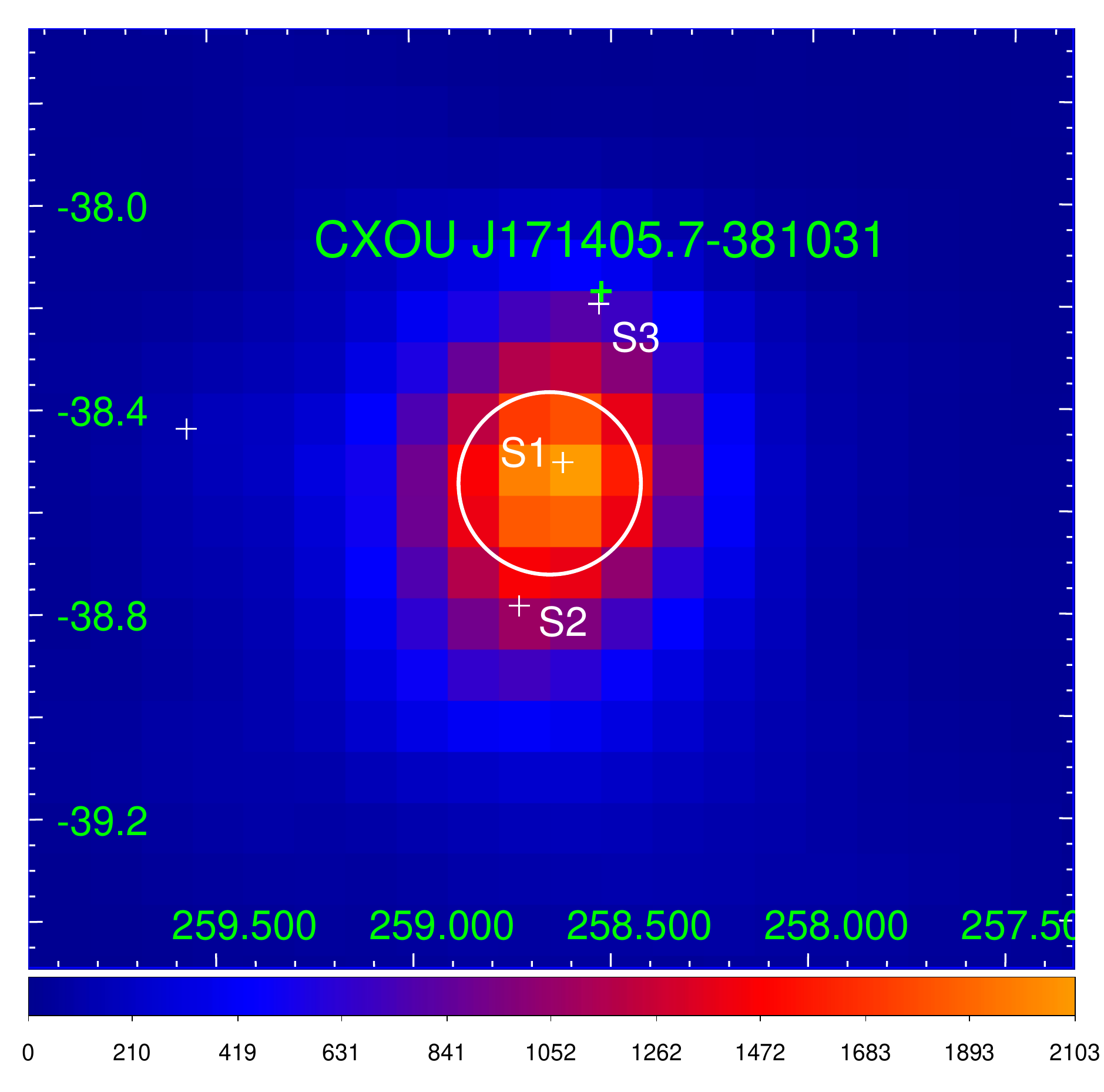}
\includegraphics[scale=0.27]{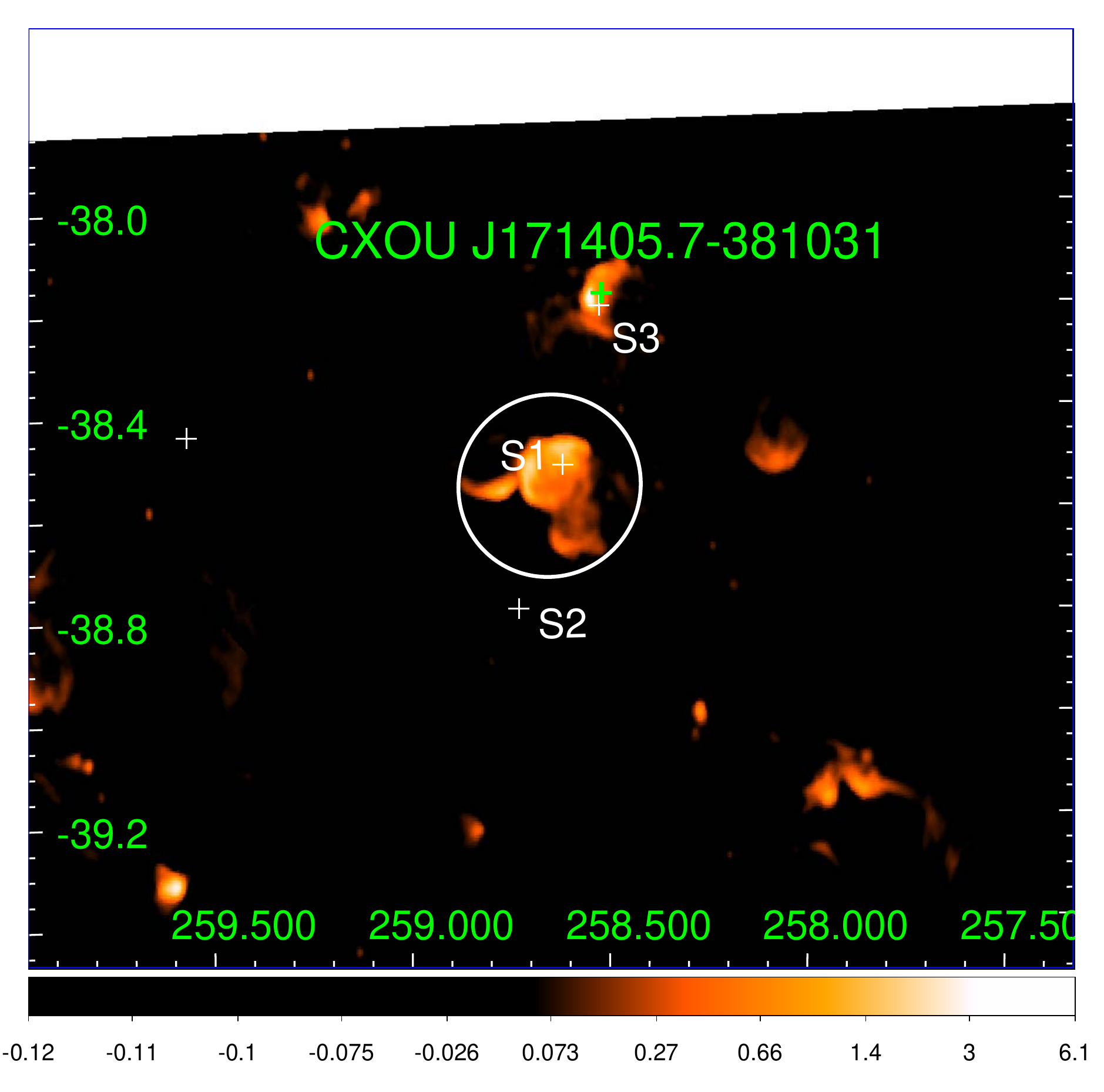}
\includegraphics[scale=0.27]{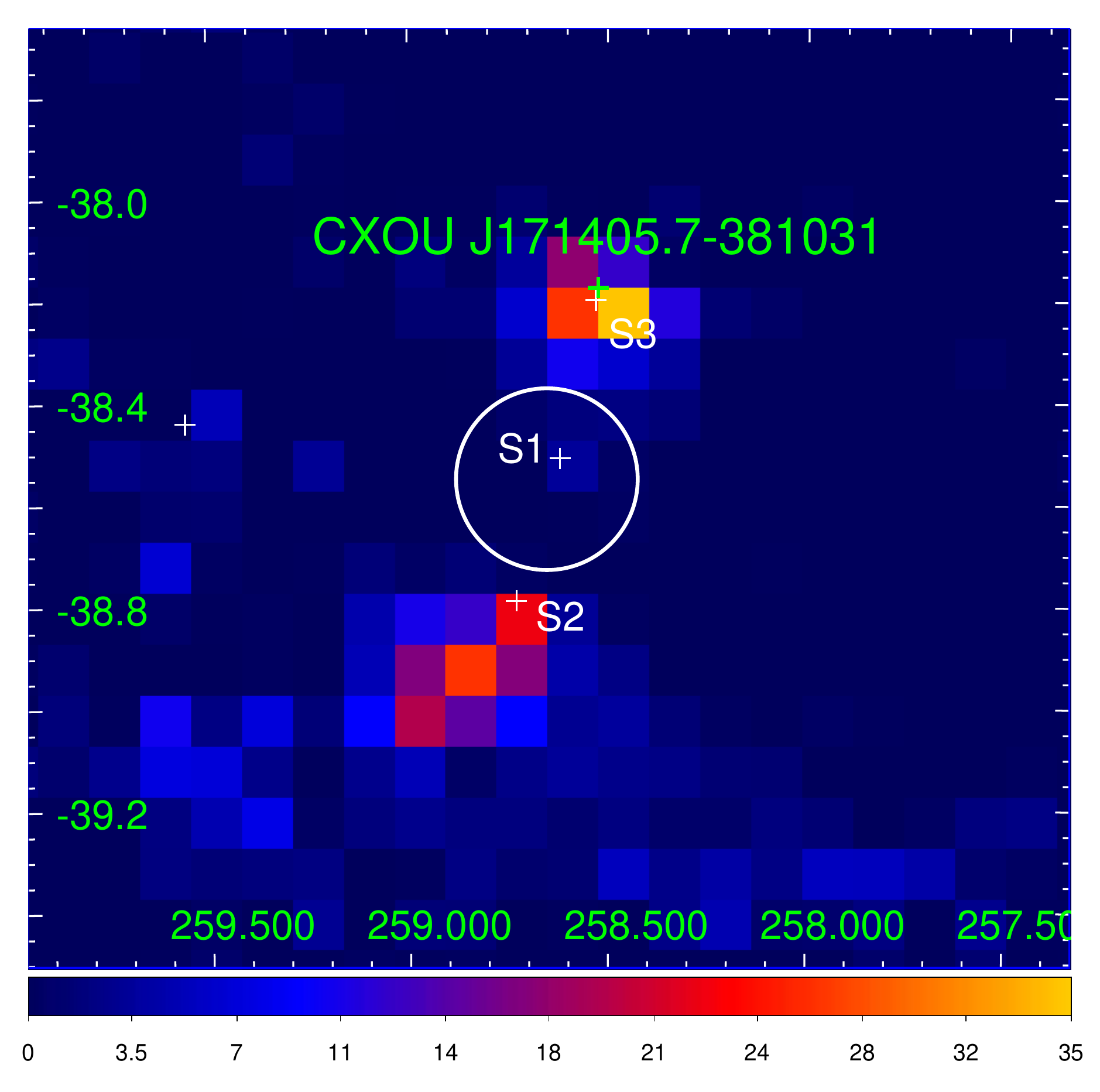}}}
\caption{\fermi-LAT fields of SGR\,1627$-$41 and CXOU\,J171405.7$-$381031.
The left column shows the relative TS maps in 0.1-300 GeV.
The middle columns are the radio map of the same region from MGPS-2.
The right column shows the residual TS maps after modelling the gamma-ray emission from SNR G337.0$-$00.1 and SNR CTB 37A, respectively.
The X-axis and Y-axis are {R.A. and decl.} referenced to J2000.
The magnetars are shown with a green cross while {other sources from the internal 6-years source list} are shown with a white cross.
The best fitted disk template are shown with a white circle.
Each source is zoomed on at the same scale as in Figure \ref{1E1841} and the size of disks can be directly compared.
The spatial models used to produce TS maps are described in the text.}
\label{SGR1627}
\label{CXOUJ1714}
\end{figure*}
\end{center}
%%%%%%%%%%%%%%%%%%%%%%%%%%%%%%%%%%

\subsection{Magnetars with associated SNRs}

\subsubsection{1E\,1841$-$045}
SNR Kes 73 is associated with 1E\,1841$-$045, which was reported as an extended gamma-ray source with a disk radius of {0$\fdg$70$ \pm$0$\fdg$03} and TS$_{ext}=$19 by Acero et al. (2016a).
A gamma-ray point source from the {internal 6-years source list} is located only 0.16$\degree$ away from 1E\,1841$-$045 {(Figure \ref{1E1841})},
which was excluded from the spatial model.
The \textit{pointlike} TS map of 1E\,1841$-$045 showed an extended source peaked near the magnetar (Figure \ref{1E1841}, top left panel).
The source is not present in the third \emph{Fermi} Source Catalog (Acero et al. 2015a, 3FGL hereafter) but it is spatially coincident {the source from the internal 6-years source list mentioned above}.
We estimated the extension of the gamma-ray source.
An extended disk with power law spectral shape has been fitted to the data. A disk of radius 0$\fdg$32$\pm$0$\fdg$03 located at R.A.$=$280$\fdg$21$\pm$0$\fdg$03, decl.$=-$4$\fdg$89$\pm$0$\fdg$03 is favored over a point-like source with TS$_{ext}$=54.4.
The gamma-ray source is significantly detected as an extended source with an overall TS value of 323.2.
The spectral parameters are reported in Table \ref{table2}.
Thus, the extended gamma-ray source detected is most likely the GeV counterpart of SNR Kes 73.
The TS map as well as the radio map from the VLA Galactic Plane Survey (VGPS, Stil et al. 2006) are shown in Figure \ref{1E1841}, left and middle panel of the corresponding row.
SNR Kes 73 is clearly seen as a {shell-like} source in the radio map around 1E\,1841$-$045 (Figure \ref{1E1841}, middle panel).
The gamma-ray extension detected is larger than the radio size of SNR Kes 73 and there is a radio complex which plausibly also contributes to the extended emission, leading to a larger radius than the radio shell.
However, the extended gamma-ray emission peaks at SNRs Kes 73 and we propose that it is the main contributor.

In order to search for gamma-ray emission from 1E\,1841$-$045, we need to exclude SNR Kes 73.
We modelled SNR Kes 73 as an extended source with spatial parameters fixed at the above values and performed the likelihood analysis.
1E\,1841$-$045 is not detected and the TS map is shown in Figure \ref{1E1841}.
By assuming a power law spectral model, we derived 95\% flux upper limits in the 0.1--1 GeV, 1--10 GeV and 0.1--10 GeV energy bands.
The results are reported in Table 1.

\subsubsection{1E\,2259$+$586}

In 3FGL, a bright gamma-ray source 3FGL J2301.2+5853 is located only 0$\fdg$02 away from 1E\,2259$+$586.
In the {internal 6-years source list, two bright gamma-ray point sources are located near 1E\,2259$+$586}, which are both positionally {coincident} with SNR CTB 109 (Figure \ref{1E2259}, middle panel).

For a detailed analysis of the 1E\,2259$+$586 region, we excluded {the two sources} from the spatial model.
The \textit{pointlike} TS map of 1E\,2259$+$586 showed a source spatially {coincident} with the magnetar (Figure \ref{1E2259}, left panel).
To check for the possible extension of the gamma-ray source, we performed the likelihood analysis in a similar way as was done for 1E\,1841$-$045.
A disk of radius {0$\fdg$20}$\pm$0$\fdg$02 located at R.A.$=${345$\fdg$46}$\pm$0$\fdg$02, decl.$=$58$\fdg$88$\pm$0$\fdg$02 is favored over a point-like source with TS$_{ext}$={23.0} (Figure \ref{1E2259}, left panel).
The gamma-ray source is significantly detected as an extended source with an overall TS value of {74.5}.
The spectral parameters are reported in Table \ref{table2}.
SNR CTB 109 is associated with 1E\,2259$+$586, which is reported as a point-like source by Acero et al. (2016a) and an extended source by Castro et al. (2012) with a consistent extension.
The TS map as well as the radio map from the Canadian Galactic Plane Survey (CGPS, Taylor et al. 2003) of the 1E\,2259$+$586 region are shown in Figure \ref{1E2259}, middle panel.
SNR CTB 109 is clearly seen as a {shell-like} structure in the radio map around 1E\,2259$+$586.
The disk radius derived from the gamma-ray data is consistent with the radio size of SNR CTB 109.
Thus, the extended gamma-ray source detected around 1E\,2259$+$586 is most likely the GeV counterpart of SNR CTB 109.

After we modelled SNR CTB 109 as an extended source with spatial parameters fixed at the above values, 1E\,2259$+$586 is not detected (Figure \ref{1E2259}, right panel).
The 95\% flux upper limits in the 0.1--1 GeV, 1--10 GeV and 0.1--10 GeV energy bands are reported in Table 1.

%%%%%%%%%%%%%%%%%%%%%%%%%%%%%%%%%%%

\begin{center}
\begin{figure*}
\centering
\vbox{
\hbox{
\includegraphics[scale=0.27]{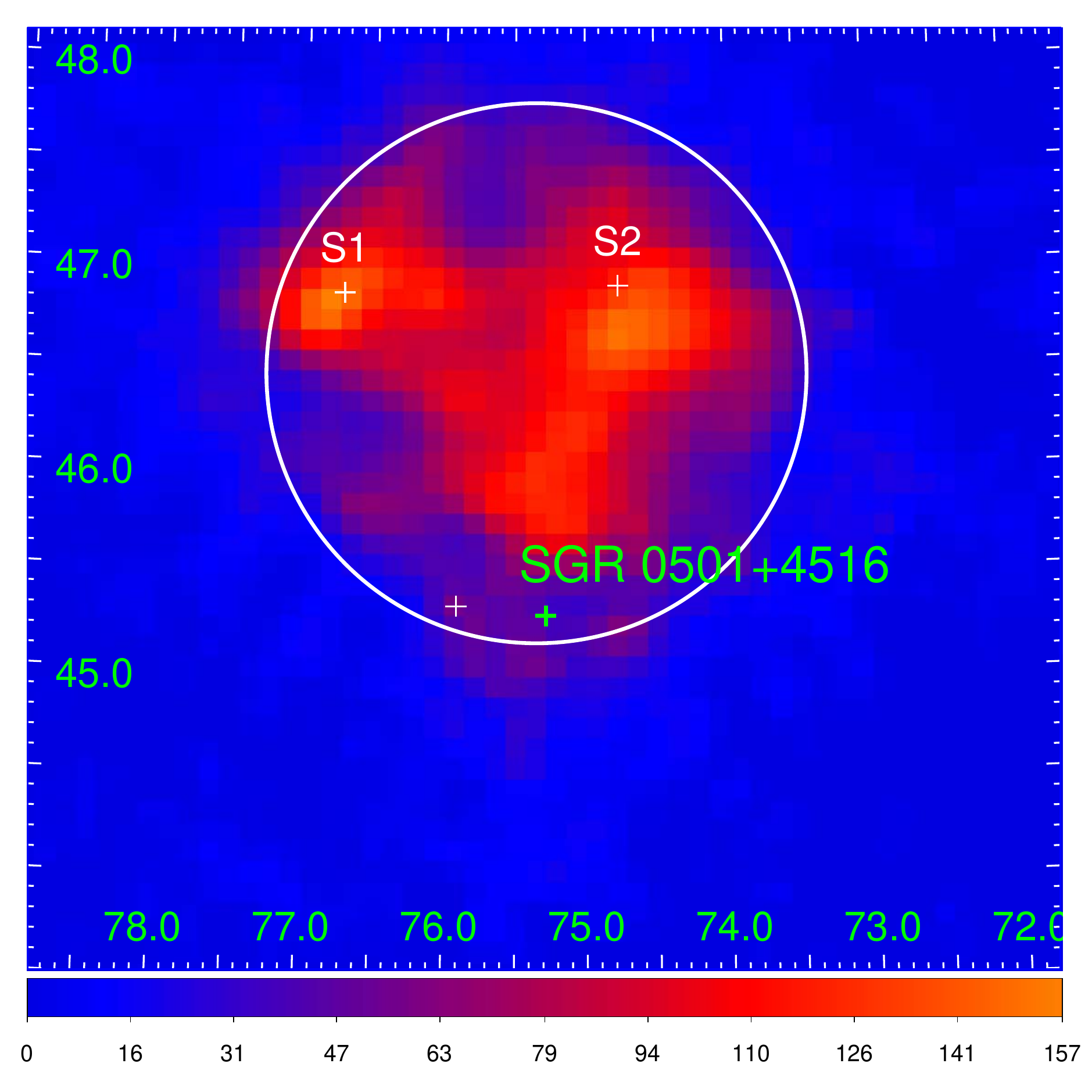}
\includegraphics[scale=0.27]{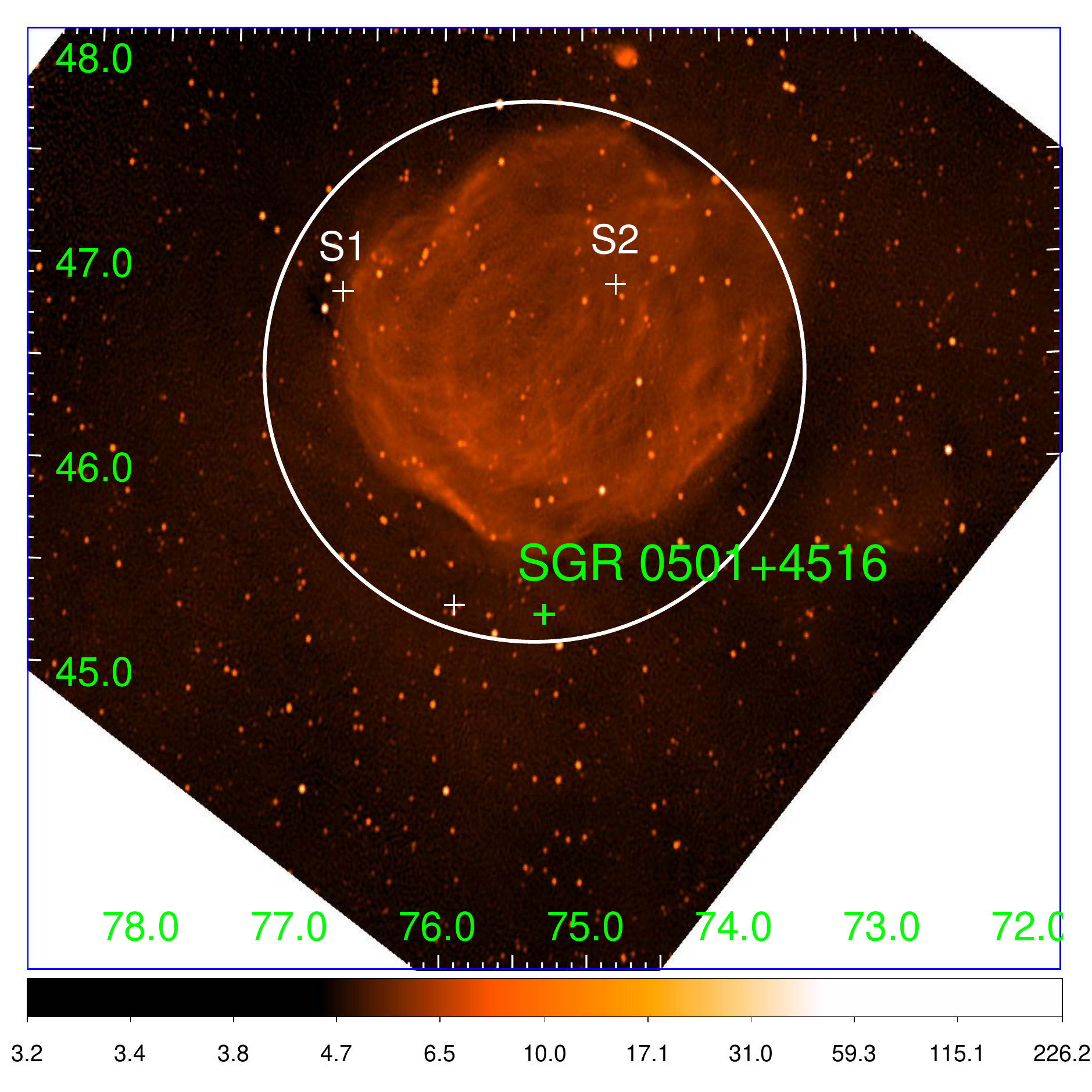}
\includegraphics[scale=0.27]{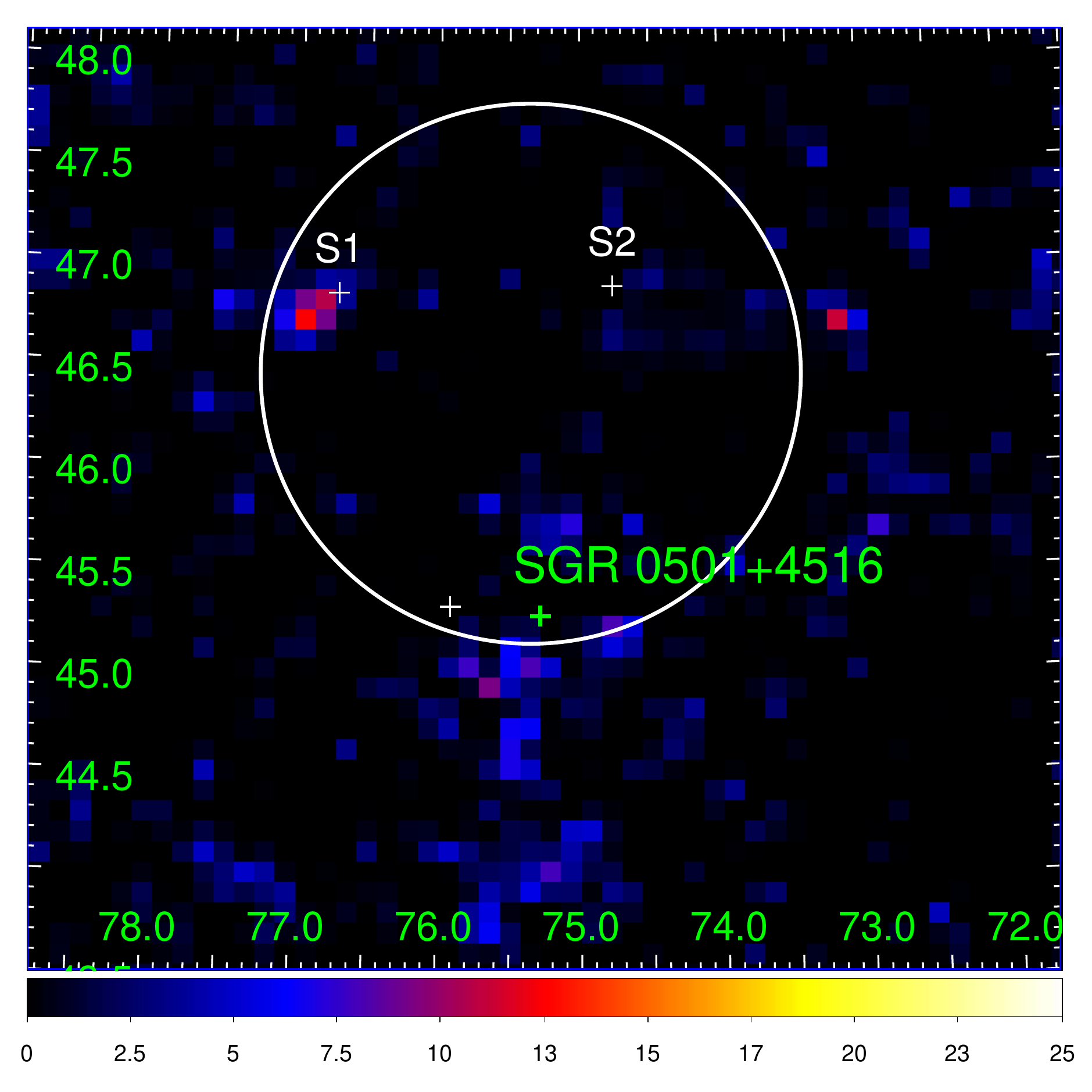}
}
\hbox{
\includegraphics[scale=0.27]{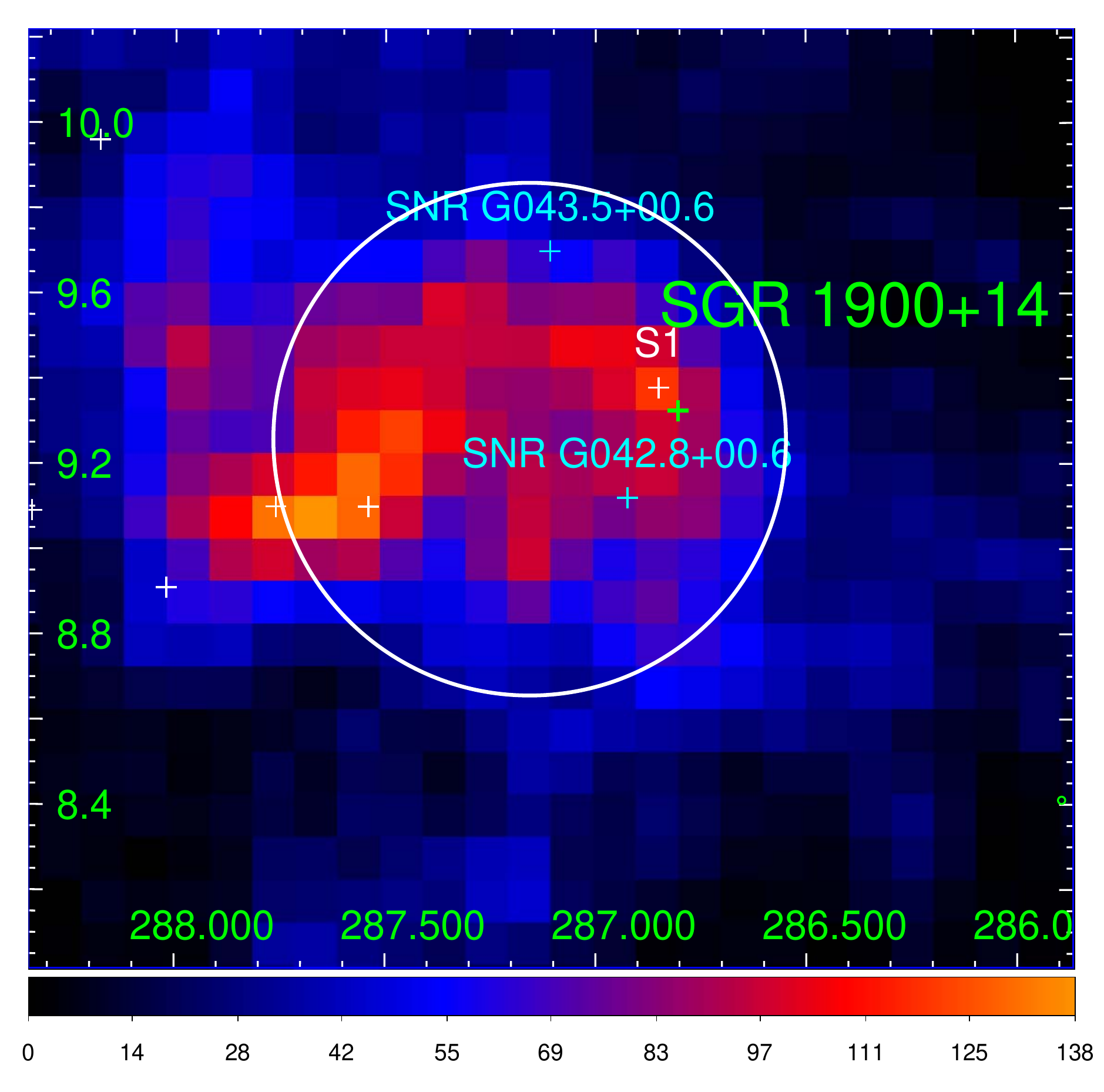}
\includegraphics[scale=0.27]{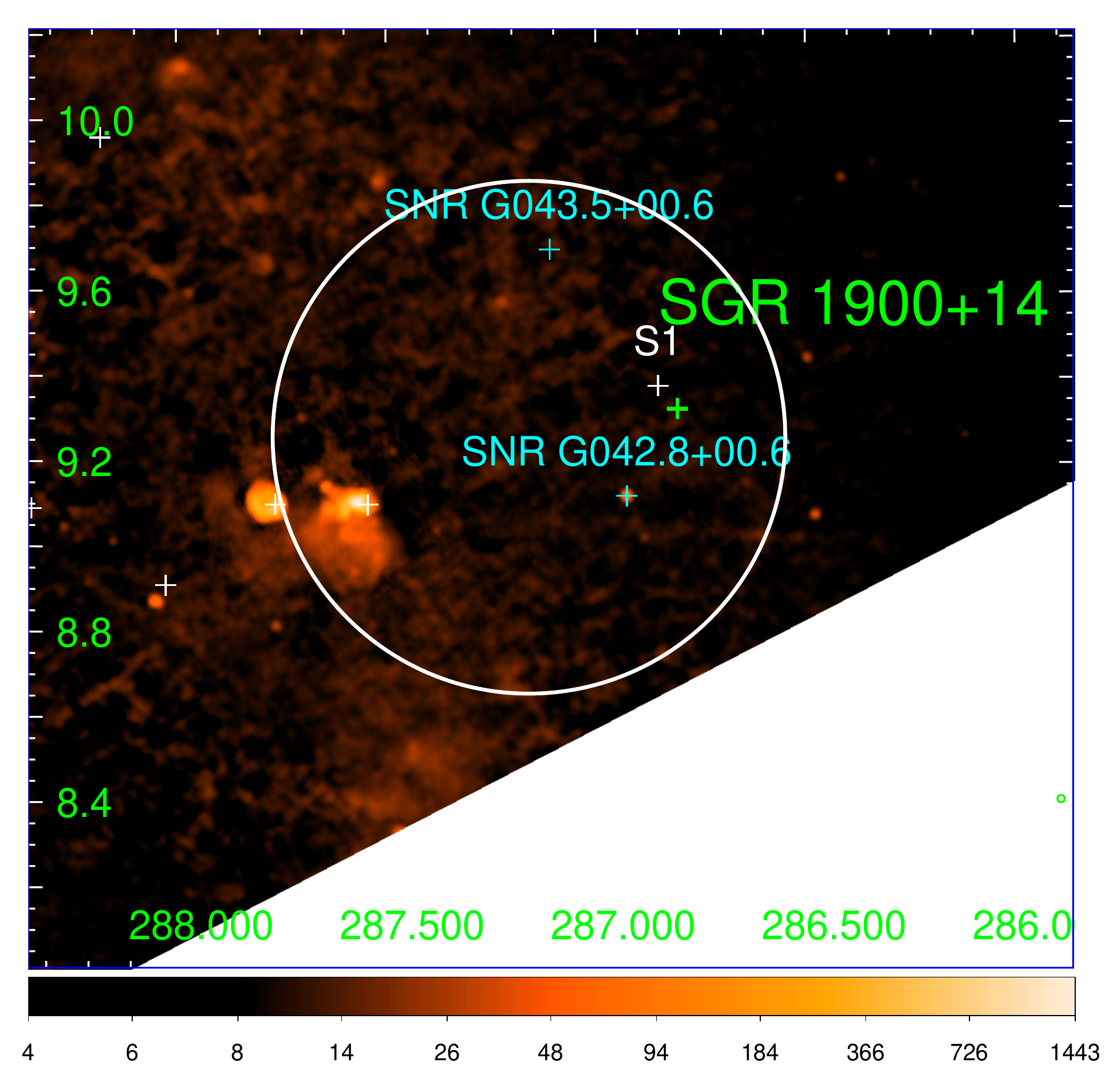}
\includegraphics[scale=0.27]{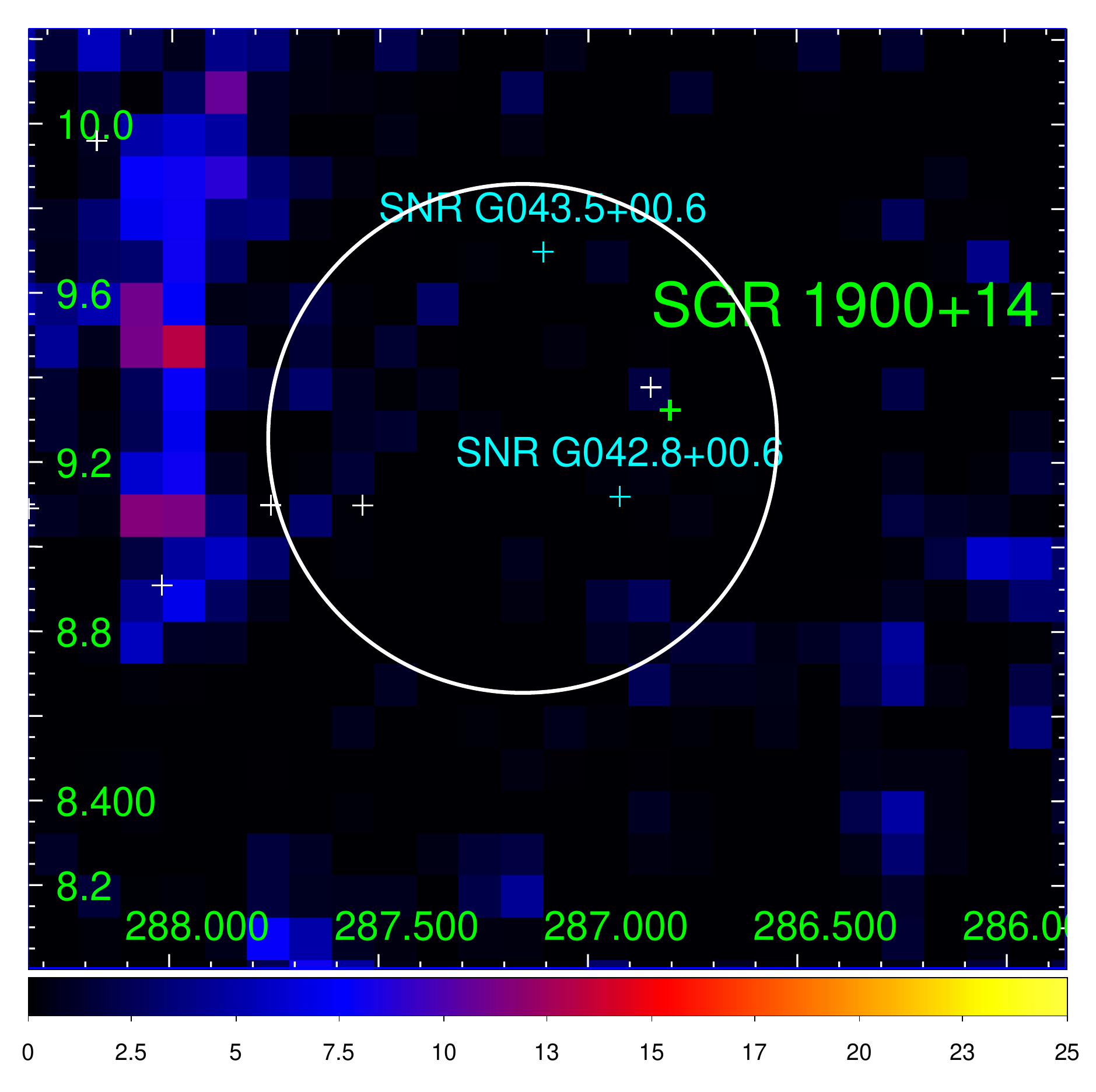}
}
\hbox{
\includegraphics[scale=0.27]{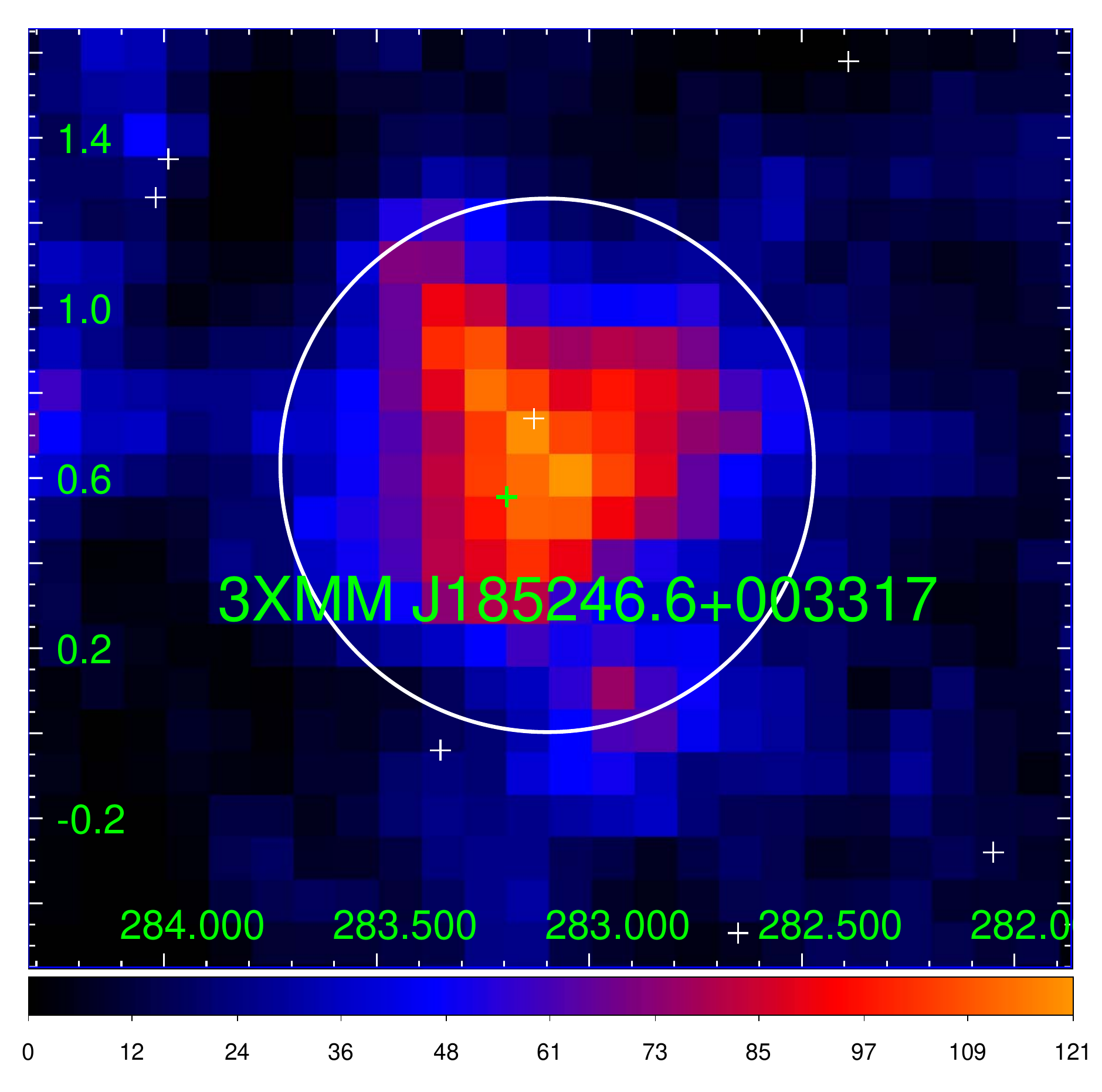}
\includegraphics[scale=0.27]{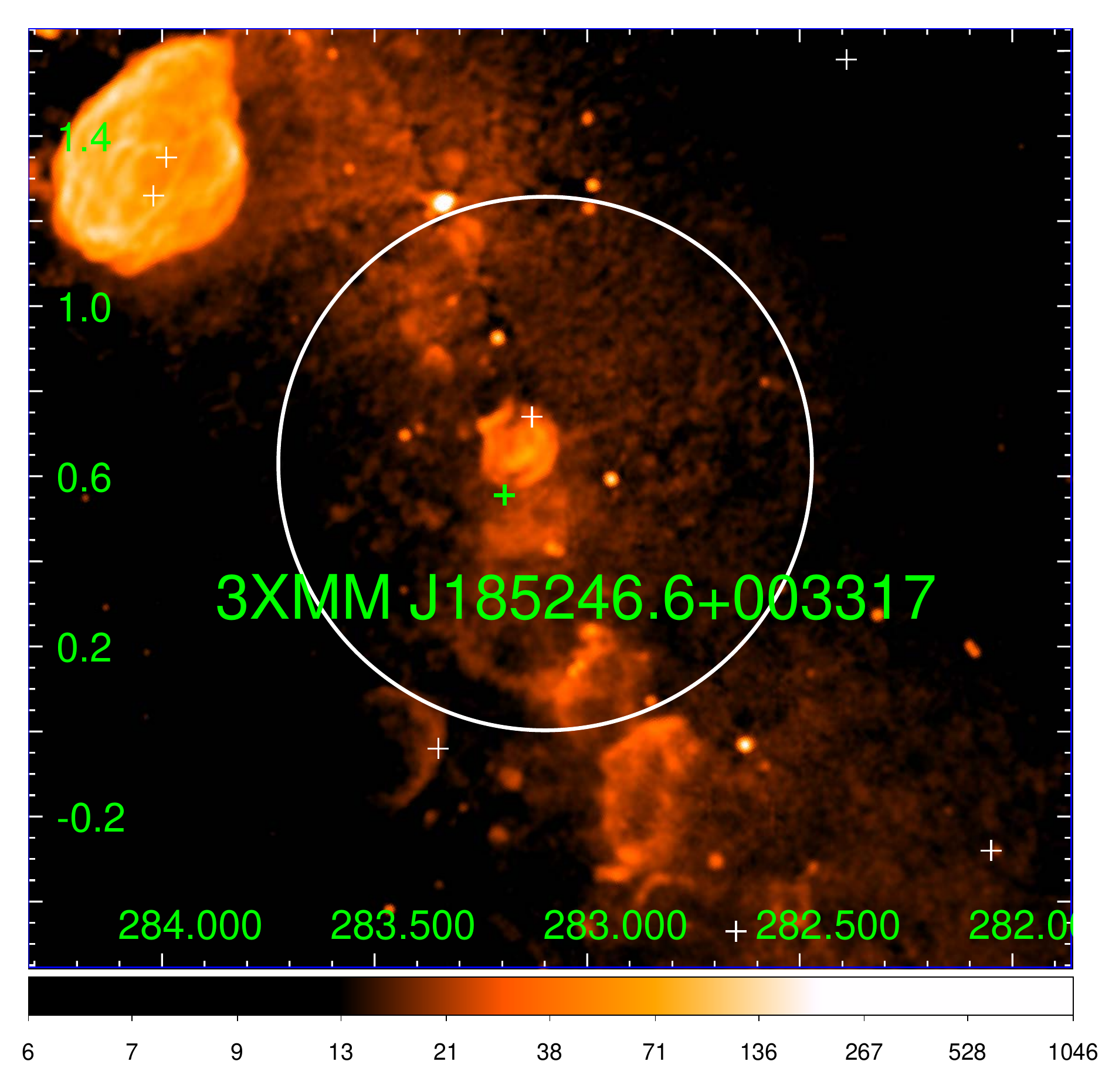}
\includegraphics[scale=0.27]{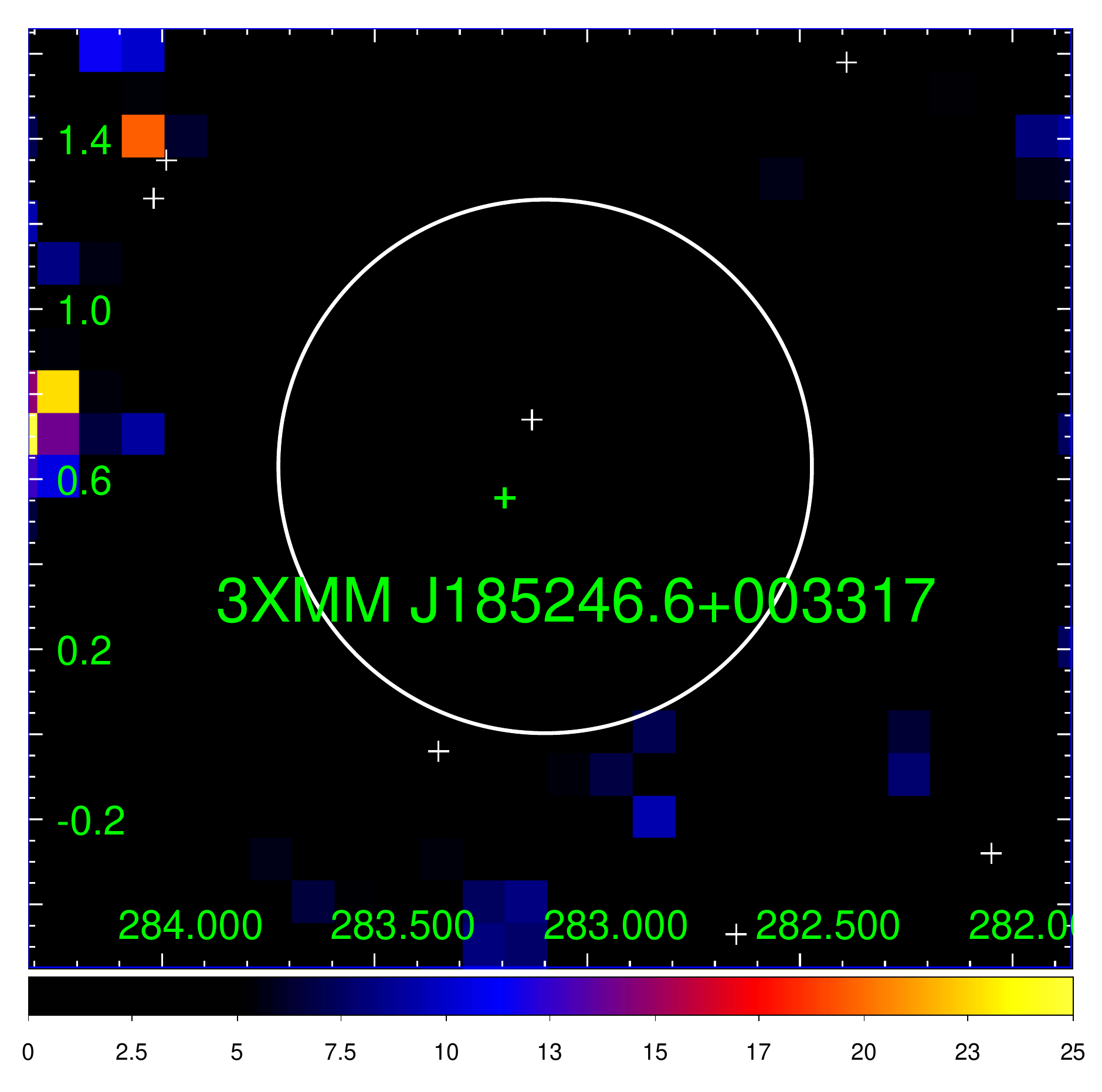}
}
}
\caption{\fermi-LAT fields of SGR\,0501$+$4516, SGR\,1900$+$14 and 3XMM\,J185246.6$+$003317.
The left column shows the relative TS maps in 0.1-300 GeV.
The middle columns are the radio map of the same region from VGPS.
The right column shows the residual TS maps after modelling the gamma-ray emission from SNR HB9, extended emission around SGR\,1900$+$14 and SNR Kes 79, respectively.
The X-axis and Y-axis are {R.A. and decl.} referenced to J2000.
The magnetars are shown with a green cross while sources in the {internal 6-years source list} are shown with a white cross.
The best fitted disk template are shown with a white circle.
SGR\,1900$+$14 and 3XMM\,J185246.6$+$003317 are zoomed on at the same scale and can be directly compared.
However, SGR\,0501$+$4516 is zoomed on at a different scale from the other two sources and cannot be directly compared.
The spatial models used to produce TS maps are described in the text.}
\label{SGR0501}
\label{SGR1900}
\label{3XMM}
\end{figure*}
\end{center}

%%%%%%%%%%%%%%%%%%%%%%%%%%%%%%%%%%

\subsubsection{Swift\,J1834.9$-$0846 \& SGR\,1833$-$0832}

Swift\,J1834.9$-$0846 is associated with SNR W41 and is in the same sky region as SGR\,1833$-$0832.
As a result, the likelihood analysis for the two sources are carried out together.
SNR W41 was detected by MAGIC and H.E.S.S. at TeV energies (Aharonian et al. 2005; Albert et al. 2006) and GeV emission is expected.
The point-like gamma-ray source 3FGL J1834.5$-$0841 from 3FGL and {two point-like gamma-ray sources from the internal 6-years source list} are located in the vicinity of the two magnetars, and are positionally {coincident} with SNR W41 (Figure \ref{Swift1834}, middle panel).
SNR W41 was once reported as an extended gamma-ray source with a disk radius of 0$\fdg$33 by Acero et al. (2016a), and by Castro et al. (2013).
Using a two-dimensional Gaussian profile, the extension of SNR W41 was reported as $\sigma$=0$\fdg$15 by Abramowski et al. (2015).
For a detailed analysis of Swift\,J1834.9$-$0846 and SGR\,1833$-$0832 as well as exploring the possible gamma-ray extension of  SNR W41, we excluded the {two sources in the internal 6-years source list} from the spatial model.
The likelihood analysis of this region showed an extended gamma-ray source spatially associated with Swift\,J1834.9$-$0846 and SGR\,1833$-$0832 (Figure \ref{Swift1834}).
To check for its possible extension, we performed the likelihood analysis in a similar way as done for 1E\,1841$-$045.
A disk of radius 0$\fdg$31$\pm$0$\fdg$02 located at R.A.$=$ 278$\fdg$64$\pm$0$\fdg$02, decl.$=$$-$8$\fdg$72$\pm$0$\fdg$02 is favored over a point-like source with TS$_{ext}$=180.4.
The gamma-ray source is significantly detected as an extended source with an overall TS value of 508.6.
The spectral parameters are reported in Table \ref{table2}.
The spectral parameters are consistent with Abramowski et al. (2015) and the extension is about twice the $\sigma$ in Abramowski et al. (2015), which are consistent considering the different extension model used--uniform disk model and two-dimensional Gaussian profile (Lande et al. 2012).
The TS map as well as the radio map from VGPS around Swift\,J1834.9$-$0846 and SGR\,1833$-$0832 are shown in Figure \ref{Swift1834}, left and middle panels.
The {shell-like} structure of SNR W41, which is associated with Swift\,J1834.9$-$0846, is apparent in Figure \ref{Swift1834}, middle panel.
No radio source is associated with SGR\,1833$-$0832.
The gamma-ray source detected is most likely the GeV counterpart of SNR W41.

We modelled SNR W41 as an extended source with spatial parameters fixed at the above values.
Swift\,J1834.9$-$0846 and SGR\,1833$-$0832 are not detected (Figure \ref{Swift1834}, right panel).
The 95\% flux upper limits in the 0.1--1 GeV, 1--10 GeV and 0.1--10 GeV energy bands are reported in Table 1.

\subsubsection{SGR\,1627$-$41}

SGR\,1627$-$41 is associated with SNR G337.0$-$00.1.
The SNR was detected as an extended gamma-ray source with a disk radius of {0$\fdg$29$\pm$0$\fdg$03} and TS$_{ext}=$43 by Acero et al. (2016a), and by Castro et al. (2013).
The point-like gamma-ray source 3FGL J1636.2$-$4734 from 3FGL, and two point-like gamma-ray sources from the internal 6-years source list are all in the vicinity of SGR\,1627$-$41 and are positionally coincident with SNR G337.0$-$00.1 (Figure \ref{SGR1627}. middle panel).
For a detailed analysis of SGR\,1627$-$41 as well as exploring the possible gamma-ray extension of SNR G337.0$-$00.1, we excluded the two sources in the internal 6-years source list from the spatial model.
The likelihood analysis of this region showed an extended gamma-ray source positionally {coincident} with SGR\,1627$-$41.
To check for the possible extension of the gamma-ray source, we performed the likelihood analysis.
Since 3FGL J1636.2$-$4734 and {one of the above two sources} are described by a LogParabola model in 3FGL and the {internal 6-years source list}, we adopted an extended disk with LogParabola spectral shape to fit the data.
A disk of radius {0$\fdg$25}$\pm$0.01 located at R.A.$=$249$\fdg$00$\pm$0$\fdg$01, decl.$=$$-${47$\fdg$54}$\pm$0$\fdg$01 is favored over a point-like source with TS$_{ext}$={133.2}.
The gamma-ray source is significantly detected as an extended source with an overall TS value of {1942.2}.
The spectral parameters are reported in Table \ref{table2}.
The TS map as well as the radio map from the Molonglo Galactic Plane Survey 2nd Epoch (MGPS-2, Murphy et al., 2007) are shown in Figure \ref{SGR1627} left and middle panels.
The radio complex of SNR G337.0$-$00.1 is clearly seen in the radio map around SGR\,1627$-$41.
The gamma-ray source detected is most likely the GeV counterpart of SNR G337.0$-$00.1.

We modelled SNR G337.0$-$00.1 as an extended source with spatial parameters fixed at the above values.
SGR\,1627$-$41 is not detected (Figure \ref{SGR1627}, right panel).
The 95\% flux upper limits in the 0.1--1 GeV, 1--10 GeV and 0.1--10 GeV energy bands are reported in Table 1.

\subsubsection{CXOU\,J171405.7$-$381031}
\label{CXOU_subsection}

CXOU\,J171405.7$-$381031 is associated with SNR CTB 37B while one gamma-ray point source {from the internal 6-years source list} is located within SNR CTB 37B (S3 in Figure \ref{CXOUJ1714}, middle panel).
The gamma-ray point source 3FGL J1714.5$-$3832 from 3FGL and {one gamma-ray source from the internal 6-years source list (S1 in Figure \ref{CXOUJ1714}, middle panel), which are $\sim$ 0$\fdg$4 away from CXOU\,J171405.7$-$381031, are positionally coincident with SNR CTB 37A.
Another two point sources (S2 and S3 in Figure \ref{CXOUJ1714}) from the internal 6-years source list} are in the vicinity.
The radio map from MGPS-2 of this region is shown in Figure \ref{CXOUJ1714}, middle panel and the two SNRs are apparent in the radio map.
Although SNR CTB 37A was once reported as a point source by Acero et al. (2016a), to exclude its influence on the analysis of CXOU\,J171405.7$-$381031, we estimated its possible extension.
S1, S2 and S3 were removed from the spatial mode.
To check for the possible extension of SNR CTB 37A, we performed the likelihood analysis as for the case of SGR\,1627$-$41.
Since 3FGL J1714.5$-$3832 and {S1} were described by a LogParabola model in 3FGL and {the internal 6-years source list} respectively, we adopted an extended disk with LogParabola spectral shape to fit the data.
A disk of radius 0$\fdg$18$\pm$0$\fdg$01 located at R.A.$=$258$\fdg$65$\pm$0$\fdg$01, decl.$=$$-$38$\fdg$55$\pm$0$\fdg$01 is favored over a point-like source with TS$_{ext}$={116.3}.
The gamma-ray source is significantly detected as an extended source with an overall TS value of {2597.2}.
The spectral parameters are reported in Table \ref{table2}.

The TS map of SNR CTB 37A is shown in Figure \ref{CXOUJ1714}, left panel.
SNR CTB 37A is bright and CXOU\,J171405.7$-$381031 is not seen in the TS map (Figure \ref{CXOUJ1714}, left panel).
In order to search for gamma-ray emission from CXOU\,J171405.7$-$381031, we need to exclude the gamma-ray emission from SNR CTB 37A.
We modelled SNR CTB 37A as an extended source with spatial parameters fixed at the above values, included {S2} and CXOU\,J171405.7$-$381031 in the spatial model and performed the likelihood analysis.
In the residual map (Figure \ref{CXOUJ1714}, right panel),  CXOU\,J171405.7$-$381031 is positionally {coincident} with the peak of a gamma-ray source.
We checked for its possible extension but it is not significant (TS$_{ext}$=5.42).
By assuming a power law spectral shape, the \textit{gtlike} analysis of this gamma-ray source resulted in a TS value of {51.0} and a photon index of 1.71$\pm$0.08.
The flux level is estimated as $1.46\pm0.19 \times10^{-11}$ erg~cm$^{-2}$s$^{-1}$ in the 0.1--300 GeV range.
We explored the presence of a spectral cut-off but it is not significant (below 3 $\sigma$).
We could not further distinguish if the source is associated with SNR CTB 37B or CXOU\,J171405.7$-$381031.
By assuming a power law spectral model, we derived 95\% flux upper limits in the 0.1--1 GeV, 1--10 GeV and 0.1--10 GeV energy bands.
The results are reported in Table 1.

\subsection{Magnetars with close-by SNRs}

\subsubsection{SGR\,0501$+$4516}

SGR\,0501$+$4516 is adjacent to SNR HB9.
The radio map from VGPS of this region is shown in Figure \ref{SGR0501}, middle panel.
The {shell-like} structure of SNR HB9 is clearly seen.
The SNR was detected as an extended gamma-ray source with a disk radius of 1$\fdg$2$\pm$0$\fdg$3 and a LogParabola spectral shape by Araya (2014).
Two point-like gamma-ray sources from the internal 6-years source list are positionally {coincident} with SNR HB9 (S1 and S2 in Figure \ref{SGR0501}, middle panel).
For a detailed analysis of SGR\,0501$+$4516 as well as exploring the possible gamma-ray extension of SNR HB9, we excluded {S1 and S2} from the spatial model.
The likelihood analysis of this region showed an extended gamma-ray source positionally {coincident} with SNR HB9.
To check for the possible extension of the gamma-ray source, we performed the likelihood analysis as in the case of SGR\,1627$-$41.
We adopted an extended disk with LogParabola spectral shape to fit the data, as in Araya (2014).
A disk of radius 1$\fdg$32$\pm$0$\fdg$05 located at R.A.$=${75$\fdg$39}$\pm$0$\fdg$05, decl.$= $$-$46$\fdg$50$\pm$ 0$\fdg$06 is favored over a point-like source with TS$_{ext}$={60.4}.
The gamma-ray source is significantly detected as an extended source with an overall TS value of {294.3}.
The spectral parameters are reported in Table \ref{table2}.

We modelled SNR HB9 as an extended source with spatial parameters fixed at the above values.
SGR\,0501$+$4516 is not detected (Figure \ref{SGR0501}, right panel).
The 95\% flux upper limits in the 0.1--1 GeV, 1--10 GeV and 0.1--10 GeV energy bands are reported in Table 1.

%%%%%%%%%%%%%%%%%%%%%%%%%%%%%%%%%%%
\begin{center}
\begin{figure}
\centering
\includegraphics[scale=0.37]{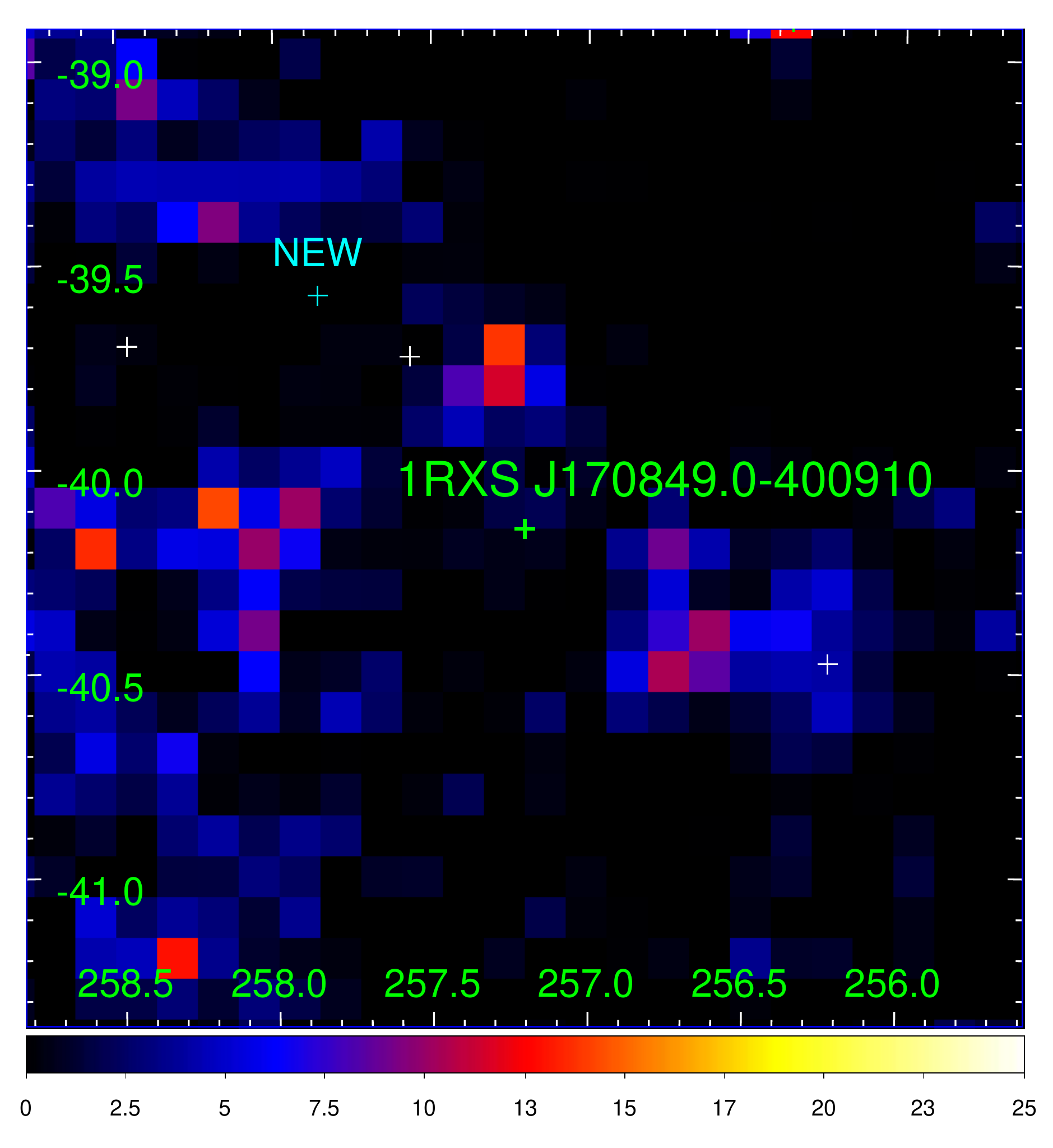}
\caption{\fermi-LAT TS maps of 1RXS\,J170849.0$-$400910 in 0.1-300 GeV with a new source added in the analysis (cyan cross).
The X-axis and Y-axis are {R.A. and decl.} referenced to J2000.
The magnetars are shown with a green cross while sources from the {internal 6-years source list} are shown with a white cross.
The spatial models used to produce TS maps are described in the text.}
\label{1RXS}
\end{figure}
\end{center}
%%%%%%%%%%%%%%%%%%%%%%%%%%%%%%%%%%

\subsubsection{SGR\,1900$+$14}

SGR\,1900$+$14 is adjacent to SNR G042.8$+$00.6 and SNR G043.5$+$00.6.
The radio map from VGPS of this region is shown in Figure \ref{SGR1900}, middle panel.
One point-like gamma-ray source from the internal 6-years source list is located only 0$\fdg$07 away from SGR\,1900$+$14 (S1 in Figure \ref{SGR1900}, middle panel).
For a detailed analysis of SGR\,1900$+$14, we excluded the source from the spatial model.
The likelihood analysis of this region showed an extended gamma-ray source positionally {coincident} with SGR\,1900$+$14, SNR G042.8$+$00.6 and SNR G043.5$+$00.6.
To check for the possible extension of the gamma-ray source, we performed the likelihood analysis in a similar way as done for 1E\,1841$-$045.
A disk of radius 0$\fdg$60$\pm$0$\fdg$06 located at R.A.$=$287$\fdg$16$\pm$0$\fdg$05, decl.$= $9$\fdg$26$\pm$0$\fdg$05 is favored over a point-like source with TS$_{ext}$={54.7}.
The gamma-ray source is significantly detected as an extended source with an overall TS value of 259.6.
The spectral parameters are reported in Table \ref{table2}. Besides SNR G042.8$+$00.6 and SNR G043.5$+$00.6, there are other radio sources positionally {coincident} with this extended gamma-ray source, which may also have contributed to the gamma-ray emission.

We modelled it as an extended source with spatial parameters fixed at the above values.
SGR\,1900$+$14 is not detected (Figure \ref{SGR1900}, right panel).
The 95\% flux upper limits in the 0.1--1 GeV, 1--10 GeV and 0.1--10 GeV energy bands are reported in Table 1.

\subsubsection{3XMM\,J185246.6$+$003317}

3XMM\,J185246.6$+$003317 is near SNR Kes 79 which is spatially coincident with one gamma-ray source from {the internal 6-years source list}.
The radio map from VGPS of this region is shown in Figure \ref{3XMM}, middle panel.
The {shell-like} structure of SNR Kes 79 is clearly seen.
The SNR was detected as a gamma-ray source by Auchettl et al. (2014).
The likelihood analysis of this region showed an extended gamma-ray source positionally {coincident} with SNR Kes 79 (Figure \ref{3XMM}, left panel).
For a detailed analysis of 3XMM\,J185246.6$+$003317 as well as exploring the possible gamma-ray extension of SNR Kes 79, we excluded {the gamma-ray source in the internal 6-years source list} from the spatial model.
To check for the possible extension of the gamma-ray source, we performed the likelihood analysis in a similar way as done for 1E\,1841$-$045.
A disk of radius 0$\fdg$63$\pm$0$\fdg$13 located at R.A.$=$ 283$\fdg$10$\pm$0$\fdg$03, decl.$=$0$\fdg$63$\pm$0$\fdg$14 is favored over a point-like source with TS$_{ext}$={108.5}.
The gamma-ray source is significantly detected as an extended source with an overall TS value of 339.5.
The extension of the gamma-ray source is larger than the size of the radio shell of SNR Kes 79 (Figure \ref{3XMM}, middle panel), which may originate from the interaction between SNR Kes 79 and the dense molecular clouds around (Auchettl et al. 2014).
The spectral parameters are reported in Table \ref{table2}.

We modelled it as an extended source with spatial parameters fixed at the above-quoted values.
3XMM\,J185246.6$+$003317 is not detected (Figure \ref{3XMM}, right panel).
The 95\% flux upper limits in the 0.1--1 GeV, 1--10 GeV and 0.1--10 GeV energy bands are reported in Table 1.

\subsection{Magnetars in crowded regions}

\subsubsection{1RXS\,J170849.0$-$400910}
\label{RXS}
1RXS\,J170849.0$-$400910 is in a crowded region on the Galactic plane.
An additional point source modelled by a simple power law was added to the spacial model.
The best position of the additional source was determined with \emph{Pointlike} as R.A.$=$ {257$\fdg$86$\pm$0$\fdg$04}, decl.$=${-39$\fdg$58$\pm$0$\fdg$04.}
There is no significant TS excess in the TS map.
The \textit{gtlike} TS value for 1RXS\,J170849.0$-$400910 is {6.6} and no clearly resolved source was seen {coincident} with it (Figure \ref{1RXS}, right panel).
The 95\% flux upper limits in the 0.1--1 GeV, 1--10 GeV and 0.1--10 GeV energy bands are reported in Table 1.
The new source yields a TS value of 197.7, a spectral index of {$3.54\pm$0.01} and a flux level of {$3.02\pm0.17 \times$}10$^{-11}$ erg~cm$^{-2}$s$^{-1}$.

%%%%%%%%%%%%%%%%%%%%%%%%%%%%%%%%%%

%%%%%%%%%%%%%%%%%%%%%%%%%%%%%%%%%%
\begin{landscape}

\begin{table*}
\centering
\scriptsize
\label{table1}
\caption{{\em Fermi}-LAT results for the SNRs included in this paper.}
\label{table2}

\begin{tabular}{llllccccc}
\\
\hline\hline                 % inserts double horizontal lines
\\
SNR name & TS &RA (J2000) &Dec (J2000)  &  TS$_{ext}$ &Radius  &Spectral Index &Fluxes & associated/adjacent magnetars   \\
                     &       & degree     & degree      &                       & degree &                         &       &                                                \\

\\
\\
\hline
\\

CTB 109         &   {74.5}       &{345.46}$\pm$0.02  & 58.88$\pm$0.02 & {23.0}   & {0.20}$\pm$0.02 &  {1.56$\pm$0.09}  & {1.44$\pm$ 0.28} & 1E\,2259$+$586/--  \\

G042.8$+$00.6; G043.5$+$00.6 & 259.6 & 287.16$\pm$0.05  & 9.26$\pm$0.05 &54.7  & 0.60$\pm$0.06 & 2.36$\pm$0.05 & 5.69$\pm$0.53 & --/SGR\,1900$+$14\\

HB9$\star$   &  {294.3}     &   {75.39}  $\pm$ 0.05    &     {46.50}$\pm$0.06   &  {60.4}   & 1.32$\pm$0.05  &  {3.81$\pm$0.41}  & {1.80}$\pm$0.12  & --/SGR\,0501$+$4516  \\

Kes 73$\dag$           &   323.2      &280.21$\pm$0.03  &$-$4.89$\pm$0.03  & 54.4   &0.32$\pm$0.03 &  2.19$\pm$0.05 & 7.17$\pm$0.59 & 1E\,1841$-$045/-- \\

Kes 79         &339.5       & 283.10$\pm$0.03    &0.63$\pm$0.14 & 108.5 & 0.63$\pm$0.13 & 2.14$\pm$0.02 & 0.93$\pm$0.02 & --/3XMM\,J185246.6$+$003317  \\

W41            &  508.6       &278.64$\pm$0.02   &$-$8.72$\pm$0.02  & 180.4  &0.31$\pm$0.02&  1.94$\pm$0.03  &11.00$\pm$0.60  &Swift\,J1834$-$0846/SGR\,1833$-$0832  \\

G337.0$-$00.1$\star$    & {1942.2}        &{249.00$\pm$0.01} &  $-${47.54}$\pm$0.01 &{133.2}  &{0.25}$\pm$0.01&{2.22}$\pm$0.04   &{14.74$\pm$0.83} &SGR\,1627$-$41/--  \\

CTB 37A$\star$    &2597.2      &   258.65 $\pm$0.01 & $-$38.55$\pm$0.01 &   116.3 & 0.18$\pm$0.01& 2.13$\pm$0.02   &13.86$\pm$0.35 & --/CXOU J171405.7$-$381031  \\

\tablecomments{Fluxes are in units of 10$^{-11}$ erg~cm$^{-2}$s$^{-1}$ and in the 0.1--300 GeV range. $\dag$: Other radio sources near Kes 73 may have contributed to the gamma-ray emission reported here. See text for more detail.
The $\star$ indicates that the index shown here is $\alpha$ for a LogParabola model. }

\end{tabular}
\end{table*}
\end{landscape}
%%%%%%%%%%%%%%%%%%%%%%%%%%%%%%%%%%

\section{Timing analysis}
\label{timing}

Pulsar rotational phases for each gamma-ray photon that passed the selection criteria could be calculated using Tempo2 (Hobbs
et al. 2006) with the \emph{Fermi} plug-in (Ray et al. 2011).
The significance of gamma-ray pulsations was evaluated with the weighted H-test (de Jager et al. 1989; de Jager \& B$\ddot{u}$sching 2010).
However, the prerequisite is to have an available precise ephemeris for the magnetars, valid during the gamma-ray data coverage of \emph{Fermi}-LAT.
Because of the noisy timing nature of magnetars, valid ephemerides during the \emph{Fermi}-LAT data coverage were not available for all the magnetars studied.

Adopting the ephemeris measured by Dib et al. (2014) with 16 years of RXTE data, we searched for gamma-ray pulsations from 1E\,1841$-$045, 1E\,2259$+$586, 1RXS\,J170849.0$-$400910 and 4U\,0142$+$614 using H-test (Hou et al. 2014).
We searched for gamma-ray pulsations for all photons below 10 GeV, exploring different values for the minimum energy (from 100 MeV to 2 GeV in a step of 100 MeV) and radius (from 0$\fdg$1 to $2\degree$ in a step of 0$\fdg$1) of the \emph{Fermi}-LAT data.
We first considered \emph{Fermi}-LAT data which are contemporaneous with RXTE which operated until the end of 2011, and then extended the last available ephemeris to the rest of \emph{Fermi}-LAT data.
No statistically significant detection of pulsations was found.
The most relevant (yet insignificant) result is obtained for the magnetar 4U\,0142$+$614 using \emph{Fermi}-LAT data with available ephemeris coverage. In the 0.4--10 GeV energy range with a data radius of 1$\fdg$3, we obtain 3.7 $\sigma$ before and 1.7 $\sigma$ after trials.

\section{DISCUSSION}
\label{discussion}

In this paper, we searched for gamma-ray emission from magnetars using 6 years of {\em Fermi}-LAT data, which greatly extended the results of Abdo et al. (2010a).
With this larger data set, and with a total of 20 magnetars searched, we can confirm that no gamma-ray emission is detected for any of the sources studied.

Twelve out of twenty magnetars are singled out by a \textit{gtlike} analysis as having TS values below 25 (Table~1).
CXOU J171405.7$-$381031 is positionally {coincident} with a point source.
However, we could not further distinguish if the emission originates from the magnetar or from SNR CTB 37B.
For all the studied magnetars we derived the deepest upper limits to date in the 0.1--300 GeV energy range.
Gamma-ray pulsations were searched for 4 magnetars having reliable ephemerides covering most of the {\em Fermi}-LAT data span, but no periodicity was significantly observed.

By applying an outer-gap model to magnetars,  Cheng \& Zhang (2001) and Zhang \& Cheng (2002) predicted detectable gamma-ray emission from SGR 1900+14 and five others AXPs within one year of {\em Fermi}-LAT observations.
These predicted flux levels are much beyond the currently imposed upper limits.
Admittedly, these models are based on a number of assumptions and parameters (see Vigan\`o et al. (2015a) for a detailed discussion).
However,  it would be unrealistic to suppose that  these parameters (e.g., the inclination angle) should be such to secure non-detections in all cases.
This point was analyzed by Tong et al. (2010) when considering the reasons for the early non-detection of 4U 0142+61.
For this particular case, even doubling the distance and using an outer gap location at 10 stellar radii or beyond the flux would have been larger than the upper limit found (see Tong et al. 2010 for details).
Takata et al. 2013 proposed a possible scenario for GeV gamma-ray emission from magnetars, in which the magnetic energy released from crust cracking could be carried into outer magnetosphere by Alfv$\acute{e}$n waves.
In this case, the predicted gamma-ray flux for 1E\,2259$+$586 is higher than the upper limits we derived.
Vigan\`o et al. (2015b) proposed an outer-gap model for gamma-ray pulsars which follows the particle dynamics to consistently compute the emission of synchro-curvature radiation.
They have applied this model to both the phase-averaged and phase-resolved spectra of the brightest pulsars.
By fitting the model to all observed {\em Fermi}-LAT pulsar spectral data, they found a strong correlation between the accelerating electric field and the magnetic field at the light cylinder (B$_{LC}$, Vigan\`o et al. 2015c).
If the correlations they found hold up to the magnetar regime (where at the light cylinder (LC), B$_{LC}$ $\sim$ 10$^{-1}$-10$^{-2}$ G, P $\sim$ 2-10 s), the accelerating  electric field in the gap would be just too weak to provide particles energetic enough as to emit gamma rays.
Their predictions that magnetars should not emit gamma rays via synchro-curvature radiation is consistent with the observational results of this paper.

Analyzing the gamma-ray emission of the regions around magnetars, we have significantly detected 7 SNRs, four of which are believed to be associated to the spatially coincident magnetar.
We have now studied them using Pass 8 data, updating their morphology and spectral parameters with respect to the 3FGL catalog (Acero et al. 2015) and Acero et al. 2016a.
The gamma-ray morphologies and luminosities of these magnetars are in line with what is expected for remnants associated with normal radio pulsars at the same age (see also Martin et al. 2014 for an X-ray comparison).

\acknowledgments

The \textit{Fermi} LAT Collaboration acknowledges generous ongoing support
from a number of agencies and institutes that have supported both the
development and the operation of the LAT as well as scientific data analysis.
These include the National Aeronautics and Space Administration and the
Department of Energy in the United States, the Commissariat \`a l'Energie Atomique
and the Centre National de la Recherche Scientifique / Institut National de Physique
Nucl\'eaire et de Physique des Particules in France, the Agenzia Spaziale Italiana
and the Istituto Nazionale di Fisica Nucleare in Italy, the Ministry of Education,
Culture, Sports, Science and Technology (MEXT), High Energy Accelerator Research
Organization (KEK) and Japan Aerospace Exploration Agency (JAXA) in Japan, and
the K.~A.~Wallenberg Foundation, the Swedish Research Council and the
Swedish National Space Board in Sweden. Additional support for science analysis during the operations phase is gratefully acknowledged from the Istituto Nazionale di Astrofisica in Italy and the Centre National d'\'Etudes Spatiales in France.

We acknowledge the support from the grants AYA2015-71042-P, SGR 2014-1073 and the National Natural Science Foundation of
China via NSFC-11473027, NSFC-11503078, NSFC-11133002, NSFC-11103020, XTP project XDA 04060604
and the Strategic Priority Research Program ``The Emergence of Cosmological Structures" of the Chinese Academy of Sciences, Grant No. XDB09000000.
NR is further supported an NWO Vidi award A.2320.0076, and via the European COST Action MP1304 (NewCOMPSTAR).


\begin{thebibliography}{99}
\bibitem[Abdo A. et al. (2010)]{abdo2009} Abdo, A. A., Ackermann, M., Ajello, M., et al. 2010a, ApJ, 725, 73
\bibitem[Abdo A. et al. (2010)]{abdo2009}Abdo, A. A., Ackermann, M., Ajello, M., et al. 2010b, A\&A, 512, 7
\bibitem[Abdo A. et al. (2010)]{abdo2009}Abdo, A. A., Ackermann, M., Ajello, M., et al. 2010c, A\&A, 523, 46
\bibitem[Abdo A. et al. (2010)]{abdo2009}Acero, F., Ackermann, M., Ajello, M., et al. 2015, ApJS, 218, 23 (3FGL)
\bibitem[Abdo A. et al. (2010)]{abdo2009}Acero, F., Ackermann, M., Ajello, M., et al. 2016a, ApJS, 224, 8 %SNR cat
\bibitem[Abdo A., et al. (2010)]{abdo2009} Acero, F., Ackermann, M., Ajello, M., et al. 2016b, ApJS, 223, 26 % diffuse mdoel

\bibitem[Abdo A. et al. (2010)]{abdo2009} Abramowski, A., Aharonian, F., Ait Benkhali, F., et al. 2015, A\&A, 574, 27



\bibitem[Abdo A. et al. (2010)]{abdo2009} Aharonian, F., Akhperjanian, A. G., Aye, K.-M., et al. 2005, Science, 307, 1938\
\bibitem[Abdo A. et al. (2010)]{abdo2009} Albert, J., Aliu, E., Anderhub, H., et al. 2006, ApJ, 643, 53\
\bibitem[Abdo A. et al. (2010)]{abdo2009} An, H.; Archibald, R., F., Hasco$\ddot{e}$t, R., et al. 2015, ApJ, 807, 93
\bibitem[Abdo A. et al. (2009)]{abdo2009}  Araya, M, 2014, MNRAS, 1, 860

\bibitem[Abdo A. et al. (2009)]{abdo2009} Atwood, W., Albert, A., Baldini L., et al. 2013, Fermi Symposium proceedings - eConf C121028 (arXiv:1303.3514)

\bibitem[Abdo A. et al. (2010)]{abdo2009} Auchettl, K., Slane, P., Castro, D. 2014, ApJ, 783, 32


%\bibitem[Abdo A. et al. (2009)]{abdo2009}   Bruel, P., et al. 2016, in prep
\bibitem[Abdo A. et al. (2009)]{abdo2009} Cheng, K. S., \& Zhang, L. 2001, ApJ, 562, 918
\bibitem[Abdo A. et al. (2009)]{abdo2009} Caliandro, G. A., Hill, A. B., Torres, D. F., et al. 2013. MNRAS, 436, 740
\bibitem[Abdo A. et al. (2009)]{abdo2009} Castro, D., Slane, P., Elission, D. C., Patnaude, D. J., 2012, ApJ, 756, 88.
\bibitem[Abdo A. et al. (2009)]{abdo2009} Castro, D., Slane, P., Carlton, A., E.  Figueroa-Feliciano, 2013, ApJ, 774, 36.

\bibitem[Abdo A. et al. (2009)]{abdo2009} de Jager, O. C., \& B$\ddot{u}$sching, I. 2010, A\&A, 517, L9
\bibitem[Abdo A. et al. (2009)]{abdo2009} de Jager, O. C., Raubenheimer, B. C., \& Swanepoel, J. W. H. 1989, A\&A, 221, 180
\bibitem[Abdo A. et al. (2009)]{abdo2009} Dib, R. \& Kaspi, V., M., ApJ, 784, 37
\bibitem[Abdo A. et al. (2009)]{abdo2009} Duncan, R. C., \& Thompson, C., 1992, ApJ, 392, 9


\bibitem[Enoto et al.(2010)]{2010ApJ...722L.162E} Enoto, T., Nakazawa, K., Makishima, K., et al.\ 2010, \apjl, 722, L162
\bibitem[Dubus (2006)]{Helene} Helene, O., 1983, NIMPR, 212, 319
\bibitem[Abdo A. et al. (2009)]{abdo2009} Hou, X., Smith, D. A., Guillemot, L., et al., 2014, A\&A, 570, 44
%\bibitem[Abdo A. et al. (2009)]{abdo2009} Israel, G. L. et al. 2015 in prep
\bibitem[Abdo A. et al. (2009)]{abdo2009} Kerr, M. 2011, PhD thesis, Univ. Washington
\bibitem[Abdo A. et al. (2009)]{abdo2009} Kuiper, L., Hermsen, W. \& Mendez, M., 2004, ApJ, 613, 1173
\bibitem[Abdo A. et al. (2009)]{abdo2009} Lande, J., Ackermann, M., Allafort, A., et al., 2012, ApJ, 756, 5
\bibitem[Abdo A. et al. (2009)]{abdo2009} Mereghetti, S, 2008, A \& ARv, 15, 225

\bibitem[Abdo A. et al. (2009)]{abdo2009} Murphy, T., Mauch, T., Green, A., et al., 2007, MNRAS, 382, 382

\bibitem[Abdo A. et al. (2009)]{abdo2009} Olausen, S. A.\& Kaspi, V. M., 2014, ApJS, 212, 60

\bibitem[Abdo A. et al. (2009)]{abdo2009} Rea, N., Esposito, P., 2011, ASSP, 21, 247

\bibitem[Abdo A. et al. (2009)]{abdo2009} Stil, J. M., Taylor, A. R., Dickey, J. M., et al. 2006, AJ 132, 1158
\bibitem[Abdo A. et al. (2009)]{abdo2009} Takata, J., Wang, Y., Wu, E. M. H., Cheng, K. S., 2013, MNRAS, 431, 2645
\bibitem[Abdo A. et al. (2009)]{abdo2009} Taylor, A. R., Gibson, S. J., Peracaula, M., et al., 2003, AJ, 125, 3145
	
\bibitem[Abdo A. et al. (2009)]{abdo2009} Thompson, C. \& Duncan, R. C., 1993, AIPC, 280, 1085

\bibitem[Abdo A. et al. (2009)]{abdo2009} Thompson, C., Lyutikov, M., \& Kulkarni, S. R., 2002, ApJ, 574, 332
\bibitem[Abdo A. et al. (2009)]{abdo2009} Tong, H., Song, L. M., \& Xu, X., 2010, ApJ, 725, 196


%\bibitem[Abdo A. et al. (2009)]{abdo2009} Vigan\`o, D., Torres, D. F., 2015a, MNRAS, 449, 3755
\bibitem[Abdo A. et al. (2009)]{abdo2009} Vigan\`o D., Torres D. F., Hirotani K., Pessah M. E., 2015a, MNRAS, 447, 1168

\bibitem[Abdo A. et al. (2009)]{abdo2009} Vigan\`o, D., Torres, D. F., Hirotani, K., Pessah, M. E., 2015b, MNRAS, 447, 2631
\bibitem[Abdo A. et al. (2009)]{abdo2009} Vigan\`o, D., Torres, D. F., Martin, J., 2015c, MNRAS, 453, 2599.
\bibitem[Abdo A. et al. (2009)]{abdo2009} Zhang, L., \& Cheng, K. S. 2002, ApJ, 579, 716

\end{thebibliography}
\end{document}